\documentclass[reprint,pra,aps]{revtex4-1}
\usepackage[utf8]{inputenc}

\usepackage{graphicx}
\usepackage{amsmath}
\usepackage{txfonts}
\usepackage{bbm}
\usepackage{amssymb}
\usepackage{amsfonts}
\usepackage[dvipsnames]{xcolor}

\usepackage[colorlinks,linkcolor=red,urlcolor=blue,citecolor=blue]{hyperref}

\newcommand{\Tr}{\mathrm{Tr}}
\newcommand{\ket}[1]{{\left \lvert#1\right\rangle}}
\newcommand{\bra}[1]{{\left\langle#1\right\rvert}}

\begin{document}

\title{Thermodynamics of decoherence}
\author{Maria Popovic}
\affiliation{School of Physics, Trinity College Dublin, College Green, Dublin 2, Ireland}
\author{Mark T. Mitchison}
\affiliation{School of Physics, Trinity College Dublin, College Green, Dublin 2, Ireland}
\author{John Goold}
\affiliation{School of Physics, Trinity College Dublin, College Green, Dublin 2, Ireland}

\begin{abstract}

We investigate the nonequilibrium thermodynamics of pure decoherence. In a pure decoherence process, the system Hamiltonian is a constant of motion and there is no direct energy exchange between the system and its surroundings. Nevertheless, the environment's energy is not generally conserved and in this work we show that this leads to nontrivial heat dissipation as a result of decoherence alone. This heat has some very distinctive properties: it obeys an integral fluctuation relation and can be interpreted in terms of the entropy production associated with populations in the energy eigenbasis of the initial state. We show that the heat distribution for a pure decoherence process is different from the distribution of work done by the initial system-bath interaction quench. Instead, it corresponds to a mixture of work distributions of cyclical processes, each conditioned on a state of the open system.  Inspired by recent experiments on impurities in ultra-cold gases, we demonstrate our general results by studying the heat generated by the decoherence of a qubit immersed within a degenerate Fermi gas in the lowest band of a species-selective optical lattice.
\end{abstract}
\maketitle

\section{Introduction}

Any real physical system is never isolated from its environment, but inevitably interacts with it. In open quantum systems, this interaction leads to the phenomenon of decoherence~\cite{Zurek1991,Zurek1994}, which destroys the relative phase between quantum states in superposition so that they can no longer be observed to interfere. Decoherence is not only a major limiting factor for entanglement-enhanced metrology~\cite{Huelga1997} and scalable quantum computation~\cite{Shor1995,Palma1996}, but is also fundamental for the quantum measurement process and the emergence of classical reality~\cite{Joos1985,Zurek2003}. Moreover, the dynamics of decoherence can now be studied in carefully controlled experiments~\cite{Myatt2000,Ratschbacher2013,Cetina2016,Liu2018,Unden2019,Skou2021} and harnessed for nondestructive measurements using auxiliary probes~\cite{Quan2006,Bylander2011,Dorner2013,Mazzola2013,Romach2015,Norris2016,Sung2019,Mitchison2020,Mitchison2021,Adam2022}. A better understanding of the decoherence process is thus important both for modern scientific applications and for the foundations of quantum physics.

Decoherence frequently occurs as a byproduct of thermalization, where an open system equilibrates by exchanging energy with its environment. Even for small quantum systems initially far from equilibrium, it is well known that thermalization dynamics is tightly constrained by the laws of thermodynamics~\cite{Spohn1978a,Esposito2010}. In contrast, in this article we are concerned with \textit{pure decoherence}, in which the energy of the open quantum system is strictly conserved and thermalization is inhibited. At first glance, the lack of direct energy exchange between system and heat bath seems to render the thermodynamics of pure decoherence trivial, even meaningless. In the following, we will show that this is not the case: quantum dephasing noise generated by a thermal environment is generally accompanied by nontrivial heat dissipation. Indeed, we prove under generic conditions that decoherence without dissipation is equivalent to static, classical phase noise: a highly restrictive situation that does not describe most realistic environments. We also demonstrate that the corresponding heat probability distribution obeys an integral fluctuation relation, and is entirely distinct from the work distribution associated with the initial system-bath interaction quench~\cite{Silva2008,Smacchia2012,Sindona2014}.

\section{Pure decoherence}

Consider an open quantum system, $S$, and a thermal bath, $B$, which together evolve under the Hamiltonian $\hat{H} = \hat{H}_S + \hat{H}_B + \hat{H}_{SB}$, with $\hat{H}_{SB}$ describing their interaction. A pure decoherence process arises when $[\hat{H}_{SB},\hat{H}_S]=0$, so that the system Hamiltonian $\hat{H}_S$ is a constant of motion. The most general interaction that satisfies this constraint is 
\begin{equation}
\label{general_dephasing_interaction}
    \hat{H}_{SB} = \sum_{n}g_{n}\hat{\Pi}_{n}\otimes\hat{V}_{n}
\end{equation}
where $\hat{\Pi}_n=\ket{n}\bra{n}$ is a projector onto the eigenstate of $\hat{H}_S$ with eigenvalue $\varepsilon_{n}$, $\hat{V}_{n}$ is a generic operator acting on the bath, and $g_{n}$ is a coupling constant. The initial state is taken to be an uncorrelated product $\hat{\rho}(0) = \hat{\rho}_S(0) \otimes \hat{\rho}_B(0)$, where $\hat{\rho}_B(0)=e^{-\beta \hat{H}_{B}}/Z_{B}$ is a thermal state of the bath with inverse temperature and partition function $Z_{B}= \Tr[e^{-\beta\hat{H}_{B}}]$. After a time $t$, the state is given by $\hat{\rho}(t) = e^{-i \hat{H} t} \hat{\rho}(0) e^{i\hat{H}t}$ (we use units where $\hbar=1$). In the eigenbasis of $\hat{H}_S$, the reduced state $\hat{\rho}_S(t) = \Tr_B[\hat{\rho}(t)]$ has matrix elements $\hat{\rho}^{mn}_S(t) = \langle m|\hat{\rho}_S(t)|n\rangle$ given by
\begin{equation}
    \label{reduced_state_matrix_elements}
    \hat{\rho}^{mn}_S(t) = e^{-i(\varepsilon_{m} - \varepsilon_n)t} \left\langle e^{i \hat{H}_n t} e^{-i\hat{H}_m t} \right\rangle_B  \hat{\rho}^{mn}_S(0),
\end{equation}
where $\langle\bullet\rangle_B = \Tr_B[ \bullet\hat{\rho}_B(0)]$ denotes a thermal average and we defined
\begin{align}
    \label{H_0_andH_1}
  \hat{H}_n = \hat{H}_B + g_n\hat{V}_n,
\end{align}
which describes the bath dynamics conditioned on state $\ket{n}$. Eq.~\eqref{reduced_state_matrix_elements} states that the diagonal matrix elements ($m=n$) are constant~\footnote{This identifies the system energy eigenstates $\{\ket{n}\}$ as the pointer states selected by the environment.}, while the off-diagonal elements ($m\neq n$) are proportional to the overlap $\langle e^{i \hat{H}_n t} e^{-i\hat{H}_m t}\rangle_B$, which decays in time whenever $\hat{H}_m\neq \hat{H}_n$ so that the bath carries information on the open system's state.

\section{Average thermodynamics}

A nonequilibrium state transformation is generally associated with energy transfer in the form of work and heat. Following the standard two-point measurement scheme~\cite{Talkner2007,Esposito2009,Campisi2011}, the work $W$ and heat $Q$ of a quantum process are defined by a difference in energy obtained by projective measurements at the initial time, $t=0$, and final time, $t=t_f$. Specifically, $W$ pertains to measurements of the total Hamiltonian, $\hat{H}$, while $Q$ corresponds to measurements of the bath Hamiltonian, $\hat{H}_B$~\cite{Esposito2010}. In order for work and heat to be simultaneously measurable, therefore, the interaction energy must be negligible at the beginning and end of the evolution~\cite{Talkner2020}. Here, we assume that the coupling is suddenly switched on at $t=0$ and switched off at $t_f$, but other switching protocols yield similar results. Note that this assumption is fully consistent with taking $\hat{\rho}(0)$ as a product state.

The open system's internal energy does not change during a pure decoherence process, $\Delta U = \langle \hat{H}_S\rangle_{t_f} - \langle \hat{H}_S\rangle_0 = 0$, where $\langle \bullet\rangle_t = \Tr[\bullet \hat{\rho}(t)]$ is the expectation value at time $t$. Naively, therefore, one may assume that there is no heat dissipation. However, the average heat absorbed by the bath is given by
\begin{equation}
    \langle Q\rangle = \langle \hat{H}_B\rangle_{t_f} - \langle \hat{H}_B\rangle_0,
\end{equation}
which may be nonzero since $\hat{H}_B$ is not conserved in general. Indeed, the interesting dynamical features of quantum dephasing noise are typically a direct consequence of the fact that $[\hat{H},\hat{H}_B]\neq 0$, as we will show by example later. 

The first law of thermodynamics for pure decoherence reads as $\langle W\rangle = \langle Q\rangle$, and from total energy conservation we find that $\langle W\rangle = -\langle \hat{H}_{SB}\rangle_{t_f} + \langle \hat{H}_{SB}\rangle_{0}$. We see that the heat produced in a pure decoherence process originates from the work done to couple the system and bath together. This work is then dissipated into the bath as heat when the system is decoupled from it at the end of the protocol. The associated average entropy production is~\cite{Landi2021}
\begin{equation}
    \label{eq:second_law}
    \langle \Sigma \rangle  = \Delta S + \beta \langle Q\rangle.
\end{equation}
Here, $\Delta S = S[\hat{\rho}_{S}(t_f)] - S[\hat{\rho}_{S}(0)]$ is the change in the von Neumann entropy of the system, where $S[\hat{\rho}] = -\Tr [\hat{\rho}\ln \hat{\rho}]$. According to the second law of thermodynamics, $\langle \Sigma\rangle \geq 0$, with equality if the process is reversible. %We will later show that both terms in Eq.~\eqref{eq:second_law} are separately non-negative in pure decoherence processes.

% % This statement has implications of practical interest and consequences on a range of experiments exploiting decoherence protocols \textcolor{red}{(cite experiment papers and discuss practical importance of decoherence)}.

\section{Heat distribution}

Nonequilibrium dynamics in the quantum regime can give rise to significant fluctuations of thermodynamic quantities. It is therefore crucial to go beyond average values and consider the full probability distribution of the absorbed heat, defined by the difference in energy of the bath measured before and after the decoherence process.  In a protocol of this kind, the form of the probability distribution of heat $P(Q)$ is well known~\cite{Goold2014,Funo2018,Popovic2021} and can be found from the characteristic function 
\begin{align}
\label{eq:charFun}
\Theta(u) & =\int_{-\infty}^{\infty}dQ e^{iuQ} P(Q),\notag \\ & = \Tr\left[ e^{i\hat{H} t_f} e^{i u\hat{H}_{B}} e^{-i\hat{H} t_f} e^{-i u\hat{H}_{B}}\hat{\rho}(0)\right],
\end{align}
with $u$ the counting-field parameter. The fluctuations of the exchanged heat between open system and environment are quantified by the statistical moments
\begin{align}
    \langle Q^k\rangle & = \int_{-\infty}^\infty dQ \, P(Q) Q^k \label{eq:heat_moments} = (-i)^k\left.\frac{d^k}{du^k} \Theta(u)\right\rvert_{u=0}.
\end{align}

We now derive a very general equation for the probability distribution of the heat dissipated through decoherence. Plugging the general form of the interaction Hamiltonian~\eqref{general_dephasing_interaction} into Eq.~\eqref{eq:charFun}, we obtain (see Appendix~\ref{app:charfun})
\begin{equation}
\label{eq:charFun4}
    \Theta (u)= \sum_{n} p_n \Theta_n (u),
\end{equation}
where $p_n=\rho^{nn}_{S}(0)$ and we have defined
\begin{equation}
\label{eq:charFun_n}
    \Theta_n (u) = \langle e^{i \hat{H}_nt_f}e^{iu\hat{H}_{B}}e^{-i \hat{H}_nt_f}e^{-iu\hat{H}_{B}}\rangle_{B}.
\end{equation}
We recognise $\Theta_n(u)$ as the characteristic function of the work done on the bath in the following (fictitious) cyclic unitary process. The bath is initialised in the equilibrium state $\hat{\rho}_B(0)$ with Hamiltonian $\hat{H}_B$. The Hamiltonian is then suddenly perturbed as $\hat{H}_B \to \hat{H}_n = \hat{H}_B + g_n\hat{V}_n$ and the bath is allowed to evolve for a time $t_f$ before the perturbation is switched off again. The work done during this cyclic process, $w$, is determined by the difference in energy that would be obtained by projective measurements of $\hat{H}_B$ at the start and finish of the evolution. This work is distributed according to
\begin{equation}
    \label{work_dist}
    P_n(w) = \sum_{j,k} \frac{e^{-\beta E_k}}{Z_B}|\langle E_j |e^{-i\hat{H}_nt_f}|E_k\rangle|^2 \delta \left(w - (E_j - E_k)\right),
\end{equation}
where $|E_j\rangle$ is an eigenvector of $\hat{H}_B$ with eigenvalue $E_j$. One readily verifies that Eqs.~\eqref{eq:charFun_n} and \eqref{work_dist} are related by a Fourier transform. The heat probability distribution is therefore given by
\begin{equation}
    \label{heat_dist_sum}
    P(Q) = \sum_n p_n P_n(w=Q).
\end{equation}
Eqs.~\eqref{eq:charFun4}--\eqref{heat_dist_sum} are the main results of the paper. They express the dissipated heat as a sum over the work done in independent cyclic processes, each one conditioned on an eigenstate $\ket{n}$ of the open system and weighted by the probability $p_n = \langle n|\hat{\rho}_S(0)|n\rangle$ that the system is initially in that state. Accordingly, the bath dynamics is described by the unital map $\hat{\rho}_B(t)=\sum_{n}p_n e^{-i\hat{H}_nt}\hat{\rho}_B(0)e^{i\hat{H}_nt}$, with $e^{-i\hat{H}_nt}$ the physical time evolution operator for the bath conditional on the state $\ket{n}$. 

At this stage, a few remarks are in order. The first moment of Eq.~\eqref{heat_dist_sum} is $\langle Q\rangle = \sum_n p_n \langle w\rangle_n$, where $\langle w\rangle_n$ is the mean work associated to the conditional distribution $P_n$. Since the average work done in any cyclic thermodynamic process is non-negative (i.e., Thomson's statement of the second law), we immediately conclude that $\langle Q\rangle \geq 0$. This inequality can also be understood as a consequence of the integral fluctuation relation
\begin{equation}
    \label{fluctuation_theorem}
    \langle e^{-\beta Q}\rangle = 1,
\end{equation}
which follows from Eq.~\eqref{heat_dist_sum} because each $P_n(w)$ obeys the Jarzynski equality, $\int d w e^{-\beta w} P_n(w) = 1$ (the free energy change vanishes for cyclic processes)~\cite{Jarzynski1997,Talkner2007}. The relation is consistent with the results discussed in \cite{Goold2015} where the same relation was seen to hold due to the unital nature of the channel acting on the bath. We also have that $\Delta S\geq 0$ because the evolution of the system density matrix in Eq.~\eqref{reduced_state_matrix_elements} is manifestly unital~\cite{Sagawa2012}. Therefore, pure decoherence processes obey a stronger bound on entropy production than the second law, since both terms in Eq.~\eqref{eq:second_law} are separately non-negative. We also note that the system entropy change $\Delta S$ is nonzero only when the system density matrix has initial coherences, while the entropy flux $\beta \langle Q\rangle$ depends only on the populations $p_n$. Therefore, the standard expression~\eqref{eq:second_law} represents a decomposition of the entropy production into contributions from initial coherences and populations in the energy eigenbasis, respectively~\cite{Francica2019,Santos2019,Francica2020}.
 
Nevertheless, the evolution of the system coherences in time is intimately connected to the presence of non-trivial heat dissipation. Indeed, our results show that finite heat dissipation generically occurs in a pure decoherence process, unless $[\hat{H}_n,\hat{H}_B] =  0$ for all $n$. We show in Appendix~\ref{app:classical_noise} that, if $\hat{H}_B$ is nondegenerate, the latter condition implies that the decoherence process is equivalent to classical, static phase noise. This contrasts sharply with naturally occurring quantum dephasing noise, which is typically characterised by a nontrivial frequency spectrum~\cite{Clerk2010,Alvarez2011,Norris2016}. Such time-dependent noise is a consequence of dynamical fluctuations of the bath, which arise when the absorption of energy drives it out of equilibrium, as we illustrate with the following example. 

\section{Qubit decoherence in a fermionic lattice environment}

To exemplify the thermodynamics of decoherence, we now focus on a specific system comprising a qubit coupled to a noninteracting fermionic bath. Our setup is motivated by recent experiments that monitored the decoherence of two-level impurity atoms embedded in a single-component atomic gas of ultracold fermions~\cite{Cetina2016}. At low temperatures, the atoms interact via $s$-wave scattering. The fermions therefore do not interact with each other due to wavefunction antisymmetry, while the impurity-fermion interaction strength is proportional to the scattering length, which is generally different for each internal state of the impurity. For simplicity, we assume that the host Fermi gas is confined to the lowest band of a one-dimensional lattice potential, e.g., a species-selective optical lattice~\cite{LeBlanc2007}. The Hamiltonian takes the form
\begin{align}
    \hat{H}_{S}	&= \frac{\epsilon}{2}\hat{\sigma}_{z} \label{eq:H_s},\\
    \hat{H}_B &= -\frac{\Omega}{2}\sum_{j=1}^{L}\left(\hat{c}_{j}^{\dagger}\hat{c}_{j+1}+\hat{c}_{j+1}^{\dagger}\hat{c}_{j}\right)\label{eq:H_b},\\
    \hat{H}_{SB} &= g\left|1\right\rangle \left\langle 1\right|\otimes \hat{c}_1^\dagger \hat{c}_1 \label{eq:H_sb}.
\end{align}
Here, $\hat{\sigma}_{z}$ is the Pauli spin operator for the qubit, with $\epsilon$ the energy level splitting, and $\hat{c}_{j}$ ($\hat{c}_{j}^{\dagger}$) is the fermionic annihilation (creation) operator for site $j$, with anticommutation relation $\{ c_{j},c_{k}^{\dagger}\} =\delta_{jk}$. $\hat{H}_{B}$ describes the tunnelling of fermions on the lattice, where $\Omega$ is the hopping amplitude and we impose periodic boundary conditions, $\hat{c}_{L+1} \equiv \hat{c}_1$. We take the thermodynamic limit by choosing the number of lattice sites $L=500$ to be large enough so that all results are independent of $L$. We also fix the average fermion number $N=250$ to half-filling, so that the Fermi energy equals $\Omega$. Finally, $\hat{H}_{SB}$ describes a collisional energy shift for the fermion localised on site $j=1$, with coupling strength $g$ to qubit state $\ket{1}$ and vanishing coupling when the qubit is in state $\ket{0}$. In the context of ultracold atoms, this can be achieved by tuning the corresponding scattering length to the zero crossing of a Feshbach resonance~\cite{Chin2010}. 

%The resulting decoherence dynamics is very rich as it manifests the orthogonality catastrophe~\cite{Anderson1967}

\begin{figure}
    \centering
    \includegraphics[scale=0.3]{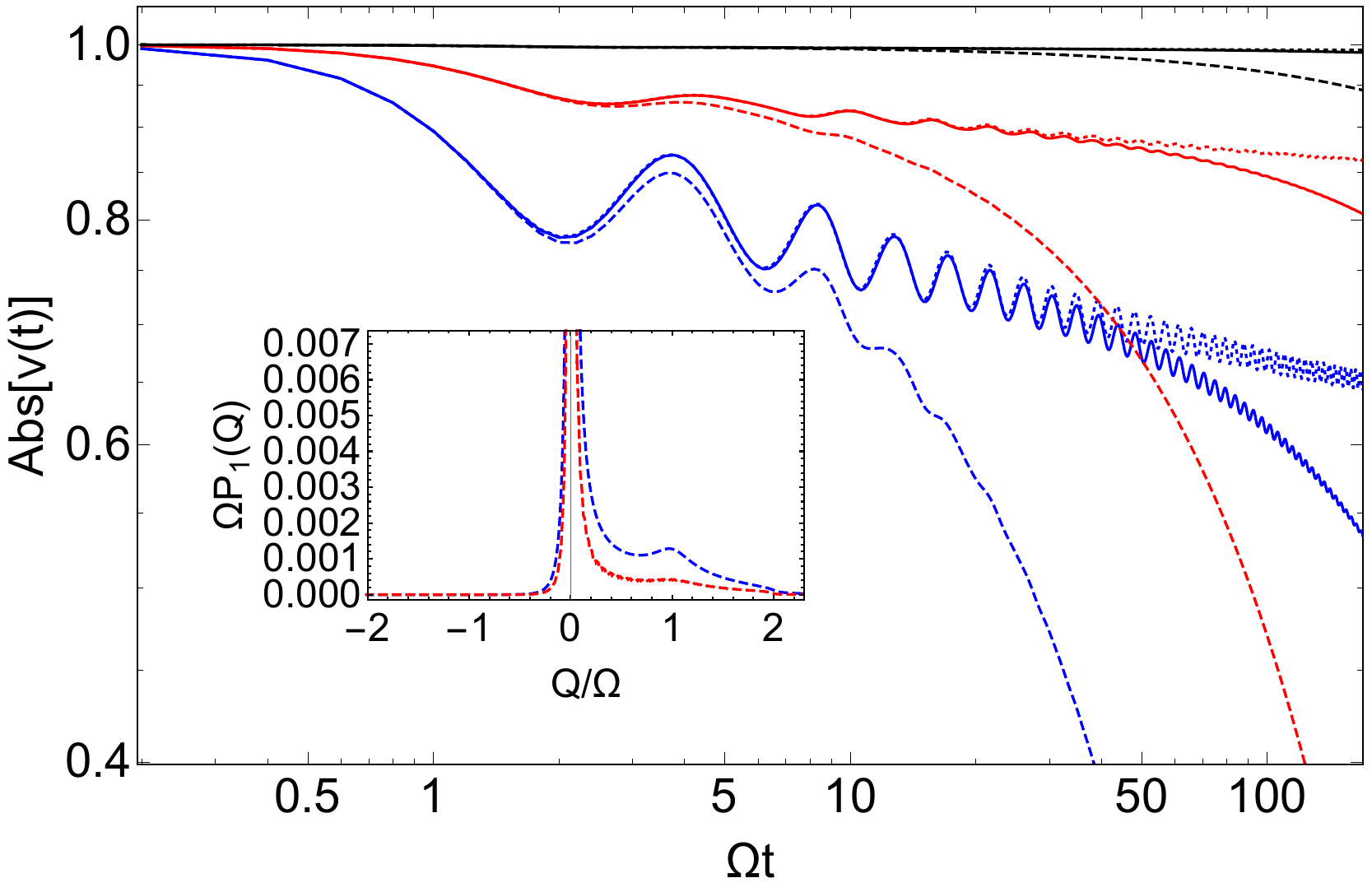}
    \caption{Decoherence functions for a qubit impurity in a 1D fermionic lattice. The value of the coupling is $g= 0.1\Omega$ (black lines), $g=0.5 \Omega$ (red lines) and $g=\Omega$ (blue lines), with temperature $T=0$ (dotted lines), $T=0.01 \Omega$ (solid lines), and $T=0.1 \Omega$ (dashed lines). Axes are in logarithmic scale. Inset: Regular part of the heat distribution $P(Q) = \tfrac{1}{2}\delta(Q)+ \tfrac{1}{2}P_{1}(Q)$ for $T=0.1 \Omega$, and $g=\Omega$ (blue, dashed), $g= 0.5\Omega$ (red, dashed).\label{fig:decoherenceFunction}}
\end{figure}

We consider a standard Ramsey interferometry protocol, following Refs.~\cite{Goold2011,Knap2012,Mitchison2020,Mitchison2021}. The bath is initialised in thermal equilibrium with the qubit in its noninteracting state $\ket{0}$. At $t=0$, a $\pi/2$-pulse prepares the qubit state $\hat{\rho}_S(0) = \left|+\right\rangle \left\langle +\right|$, with $\left|+\right\rangle = (\left|1\right\rangle+\left|0\right\rangle)/\sqrt{2}$ a superposition of the impurity eigenstates. According to Eq.~\eqref{reduced_state_matrix_elements}, the qubit coherence evolves as $\hat{\rho}^{10}_S(t) = e^{-i\epsilon t} \nu(t) \hat{\rho}^{10}_S(0)$, where $\nu(t) = \langle e^{i\hat{H}_0t} e^{-i\hat{H}_1 t}\rangle_B$ is the decoherence function with $\hat{H}_0 = \hat{H}_B$ and $\hat{H}_1 = \hat{H}_B + g \hat{c}_1^\dagger\hat{c}_1$. This complex function may be experimentally extracted by applying a second $\pi/2$-pulse with a variable phase and measuring the final qubit populations~\cite{Cetina2016,Skou2021,Adam2022}.

The decoherence function $\nu(t)$ for a bath of fermionic modes can be computed exactly through the functional determinant approach~\cite{Klich2003,Levitov2005,Schmidt2018}, which expresses the average of a product of many-body exponential operators in terms of single-particle quantities. In particular, the decoherence function is found from
\begin{equation}
\label{eq:decoherenceFunction}
\nu(t)= \det\left[1-\hat{n}+\hat{n}e^{i \hat{h}_{0} t}e^{-i\hat{h}_{1}t}\right],
\end{equation}
where $\hat{n}=\left(e^{\beta\left(\hat{h}_{0}-\mu\right)}+1\right)^{-1}$ is the single-particle occupation number operator, $\mu$ is the chemical potential and $\hat{h}_{0}$ and $\hat{h}_{1}$ are the single-particle counterparts of the Hamiltonians $\hat{H}_{0}$ and $\hat{H}_{1}$. The absolute value of $\nu(t)$ is shown in Fig.~\ref{fig:decoherenceFunction} for different temperatures and coupling constants. The decoherence function crosses over from power-law to exponential decay after the thermal timescale $\beta$. This behavior is a manifestation of the Fermi-edge singularity~\cite{Mahan1967,Nozieres1969} associated with Anderson's orthogonality catastrophe~\cite{Anderson1967}. Decoherence arises because collisions with the impurity excite particle-hole pairs in the gas, gradually reducing the overlap between the perturbed and unperturbed states of the environment. These excitations are initially restricted to the vicinity of the Fermi surface due to the Pauli exclusion principle, generating the slow power-law decoherence seen in Fig.~\ref{fig:decoherenceFunction} for $t\ll  \beta$. Thermal broadening of the Fermi surface eventually leads to the onset of exponential decay  when $t\gtrsim \beta$. At stronger coupling, we also observe oscillations of the qubit coherence (blue curves in Fig.~\ref{fig:decoherenceFunction}), indicating a highly non-Markovian (nondivisible) evolution~\cite{Breuer2016}. We therefore see that nontrivial decoherence emerges here as a direct consequence of the dissipation of energy into the fermionic bath. 

\begin{figure}
    \centering
    \includegraphics[scale=0.3, trim = 0mm 0mm 0mm 11mm, clip]{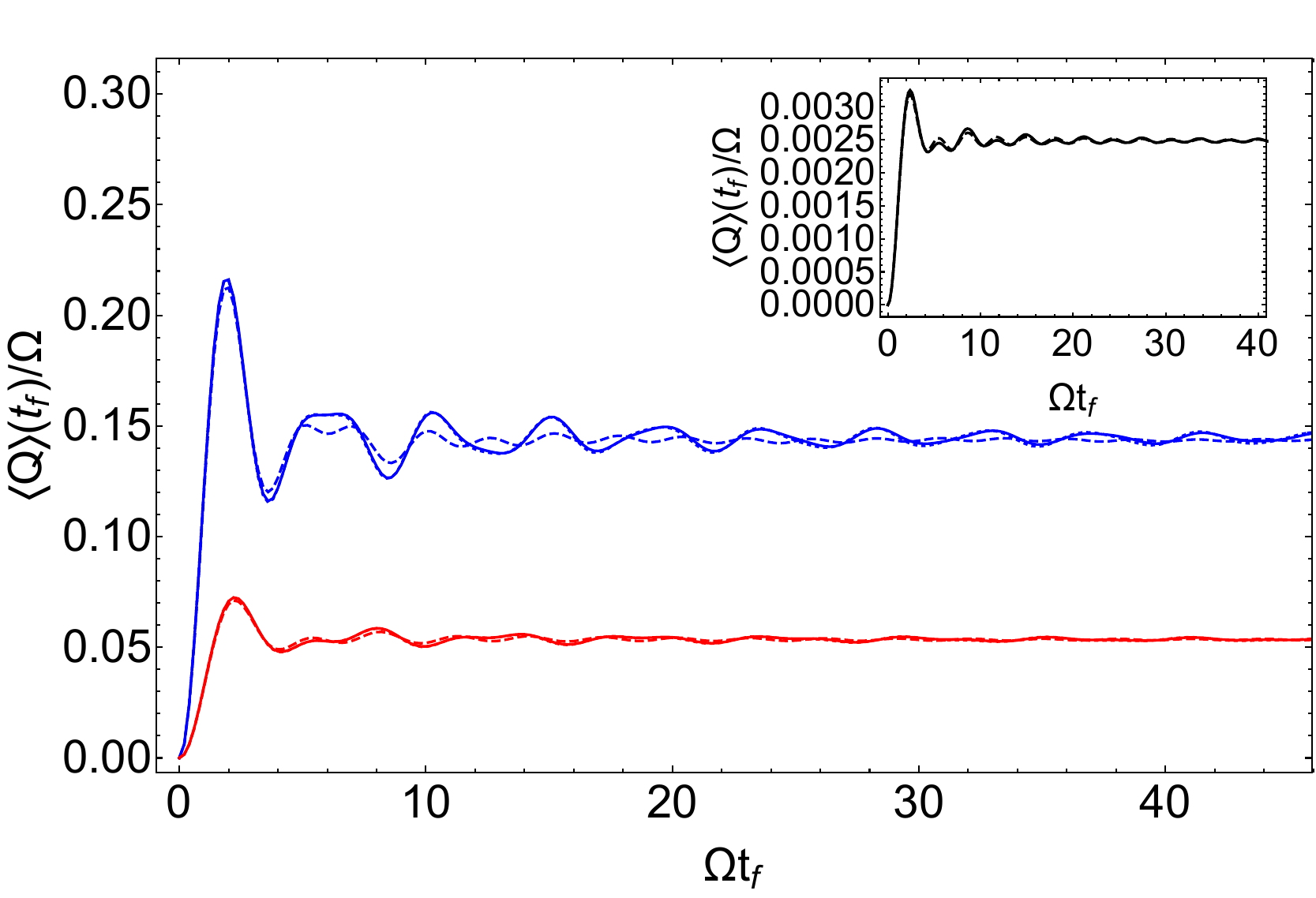}
    \caption{Heat transfer as a function of time for a qubit in a 1D fermionic lattice. The values of the coupling are $g=0.1\Omega$ (black lines), $g=0.5\Omega$ (red lines) and $g=\Omega$ (blue lines), with temperatures $T=0$ (dotted lines), $T=0.01 \Omega$ (solid lines), and $T=0.1 \Omega$ (dashed lines). \label{fig:heat Function}}
\end{figure}

The characteristic function of heat follows from Eq.~\eqref{eq:charFun4} as
\begin{equation}
    \label{eq:ThetaModel}
    \Theta (u) = \frac{1}{2}+\frac{1}{2}\left\langle e^{i t_f \hat{H}_{1}}e^{iu\hat{H}_{0}}e^{-i t_f \hat{H}_{1}}e^{-iu\hat{H}_{0}}\right\rangle _{B}.
\end{equation}
The first, constant term pertains to the $\ket{0}$ component of the qubit's initial superposition state, which leads to exactly zero heat dissipation, i.e., $P_0(Q) = \delta(Q)$. The second term in Eq.~\eqref{eq:ThetaModel} corresponds to the interacting state $\ket{1}$ and can be expressed as a functional determinant (see Appendix~\ref{app:levitov}). The probability distribution $P_{1}(Q)$ associated to this second term is shown in the inset of Fig.~\ref{fig:decoherenceFunction}, for a fixed temperature and two different values of the coupling constant. The divergence near $Q=0$ is a hallmark of the Fermi edge singularity, which is centred around zero energy as a consequence of the cyclic nature of the process. For stronger coupling, there is also a feature near the Fermi energy $Q=\Omega$, which can be attributed to the creation of particle-hole excitations at the bottom of the band~\cite{Knap2012}. 

\begin{figure}
    \centering
    \includegraphics[scale=0.3]{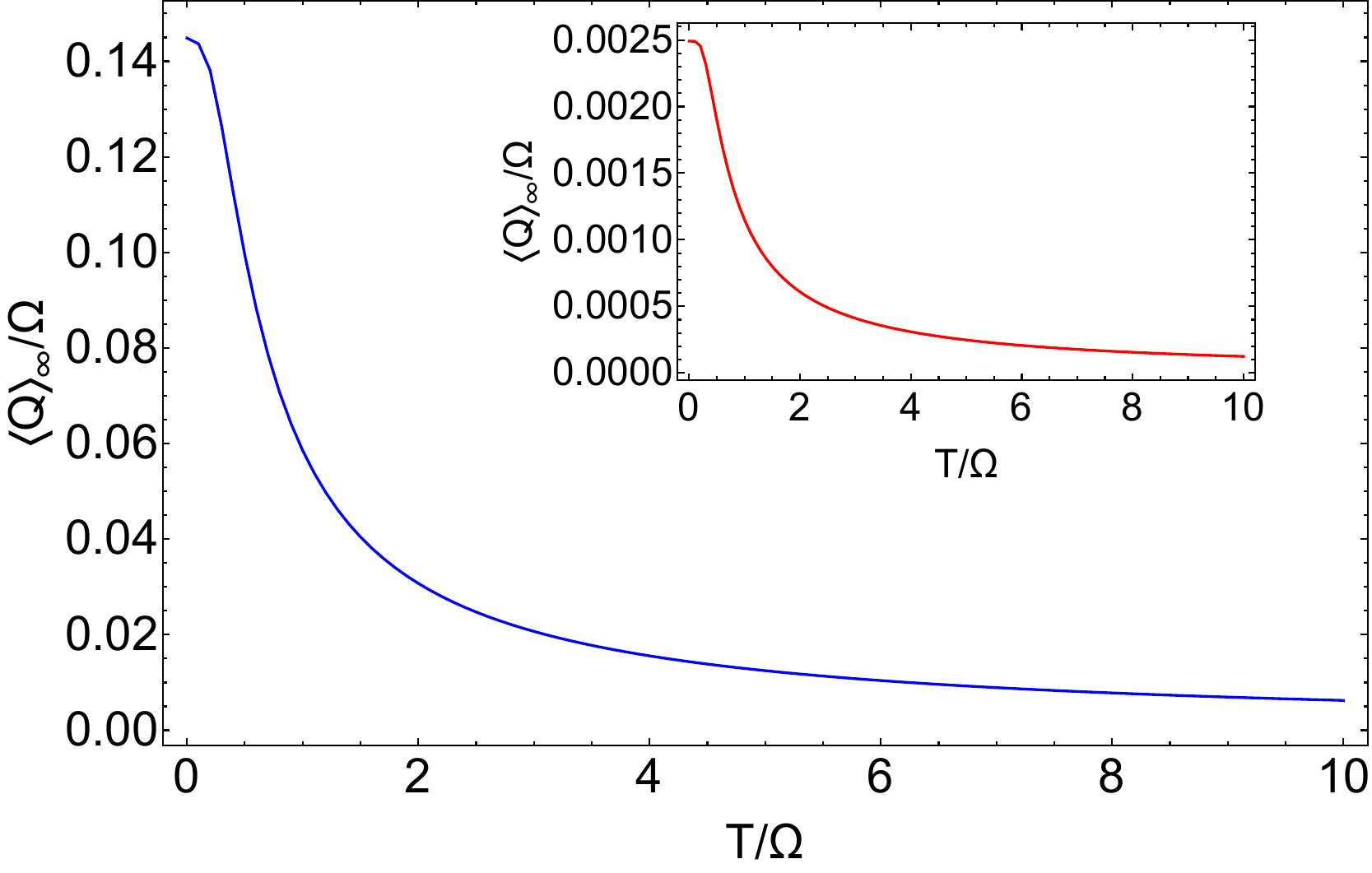}
    \caption{Long-time limit of the heat transfer as a function of temperature for a qubit in a 1D fermionic lattice, with couplings $g=\Omega$ (blue lines) and $g=0.1\Omega$ (inset, red lines). \label{fig:heatFunction_T}}
\end{figure}

The mean heat $\langle Q \rangle$ is shown in Fig.~\ref{fig:heat Function} as a function of time for the same parameters as Fig.~\ref{fig:decoherenceFunction}. We observe that $\langle Q\rangle$ grows with protocol time for short evolutions, $\Omega t_f\lesssim 1$, and then executes long-lasting oscillations around a finite value for relatively long protocols $\Omega t_f \gg 1$ before eventually settling to a constant. These intermediate-time oscillations can be understood as a consequence of the finite bandwidth of the fermionic lattice, and are more prominent for strong coupling, $g\gtrsim \Omega$. The temperature dependence of the mean heat absorbed at asymptotically long times is shown in Fig.~\ref{fig:heatFunction_T}, for two different values of the coupling constant. As its temperature increases, the bath absorbs less heat during the decoherence process, which reflects the reduced disturbance of the high-temperature Fermi sea by the impurity perturbation.

\section{Conclusions}

Decoherence and heat dissipation are commonly considered to be complementary manifestations of irreversibility in open quantum systems. Our results show that these two processes are in fact inextricably linked: the emergence of classicality through decoherence comes at a fundamental energetic cost. This heat dissipation is a subtle aspect of decoherence that, to our knowledge, has not yet been investigated in the literature on quantum thermodynamics. We note that several recent works have identified dissipation arising from changes in coherence, by considering the energetics of the open quantum system alone~\cite{deLima2020,Alipour2022,Ahmadi2023}. In contrast, by considering the full system-bath dynamics, we have shown that dissipation and pure decoherence go hand in hand even though the internal energy of the open system remains constant. Beyond their foundational implications, our findings are directly relevant for ongoing efforts to harness decoherence for nondestructive measurements of noise in quantum devices. The unavoidable dissipation of energy implies an intrinsic disturbance due to such measurements~\cite{Wang2021}, which can be quantitatively assessed using the general framework developed here. 

The experimentally relevant example of an impurity immersed in an ultra-cold Fermi gas~\cite{Cetina2016} highlights the rich physics which is unveiled from considering the thermodynamics of decoherence. However, our framework can equally well be applied to a range of other situations, such as ultracold bosonic environments where dephasing impurities have recently been realized~\cite{Skou2021,Adam2022}, as well as strongly interacting systems. We hope that our work will inspire further investigations of the peculiar thermodynamic features of decoherence and associated properties in diverse physical settings. 

\begin{acknowledgments}
We thank P.~R.~Eastham for useful discussions. This work was funded by the European Research Council Starting Grant ODYSSEY (Grant Agreement No. 758403) and the EPSRC-SFI joint project QuamNESS. J.~G. is supported by a SFI-Royal Society University Research Fellowship. M.~T.~M acknowledges support from a Royal Society-Science Foundation Ireland University Research Fellowship (URF\textbackslash R1\textbackslash 221571), and funding from the European Commission's Quantum Flagship via the Horizon Europe project ASPECTS (Grant Agreement No.~101080167).
\end{acknowledgments} 

\bibliographystyle{apsrev4-1}
\bibliography{bibliography}{}

\begin{thebibliography}{60}%
\makeatletter
\providecommand \@ifxundefined [1]{%
 \@ifx{#1\undefined}
}%
\providecommand \@ifnum [1]{%
 \ifnum #1\expandafter \@firstoftwo
 \else \expandafter \@secondoftwo
 \fi
}%
\providecommand \@ifx [1]{%
 \ifx #1\expandafter \@firstoftwo
 \else \expandafter \@secondoftwo
 \fi
}%
\providecommand \natexlab [1]{#1}%
\providecommand \enquote  [1]{``#1''}%
\providecommand \bibnamefont  [1]{#1}%
\providecommand \bibfnamefont [1]{#1}%
\providecommand \citenamefont [1]{#1}%
\providecommand \href@noop [0]{\@secondoftwo}%
\providecommand \href [0]{\begingroup \@sanitize@url \@href}%
\providecommand \@href[1]{\@@startlink{#1}\@@href}%
\providecommand \@@href[1]{\endgroup#1\@@endlink}%
\providecommand \@sanitize@url [0]{\catcode `\\12\catcode `\$12\catcode
  `\&12\catcode `\#12\catcode `\^12\catcode `\_12\catcode `\%12\relax}%
\providecommand \@@startlink[1]{}%
\providecommand \@@endlink[0]{}%
\providecommand \url  [0]{\begingroup\@sanitize@url \@url }%
\providecommand \@url [1]{\endgroup\@href {#1}{\urlprefix }}%
\providecommand \urlprefix  [0]{URL }%
\providecommand \Eprint [0]{\href }%
\providecommand \doibase [0]{http://dx.doi.org/}%
\providecommand \selectlanguage [0]{\@gobble}%
\providecommand \bibinfo  [0]{\@secondoftwo}%
\providecommand \bibfield  [0]{\@secondoftwo}%
\providecommand \translation [1]{[#1]}%
\providecommand \BibitemOpen [0]{}%
\providecommand \bibitemStop [0]{}%
\providecommand \bibitemNoStop [0]{.\EOS\space}%
\providecommand \EOS [0]{\spacefactor3000\relax}%
\providecommand \BibitemShut  [1]{\csname bibitem#1\endcsname}%
\let\auto@bib@innerbib\@empty
%</preamble>
\bibitem [{\citenamefont {Zurek}(1991)}]{Zurek1991}%
  \BibitemOpen
  \bibfield  {author} {\bibinfo {author} {\bibfnamefont {W.~H.}\ \bibnamefont
  {Zurek}},\ }\href {\doibase 10.1063/1.881293} {\bibfield  {journal} {\bibinfo
   {journal} {Physics Today}\ }\textbf {\bibinfo {volume} {44}},\ \bibinfo
  {pages} {36} (\bibinfo {year} {1991})}\BibitemShut {NoStop}%
\bibitem [{\citenamefont {Zurek}\ and\ \citenamefont {Paz}(1994)}]{Zurek1994}%
  \BibitemOpen
  \bibfield  {author} {\bibinfo {author} {\bibfnamefont {W.~H.}\ \bibnamefont
  {Zurek}}\ and\ \bibinfo {author} {\bibfnamefont {J.~P.}\ \bibnamefont
  {Paz}},\ }\href {\doibase 10.1103/PhysRevLett.72.2508} {\bibfield  {journal}
  {\bibinfo  {journal} {Phys. Rev. Lett.}\ }\textbf {\bibinfo {volume} {72}},\
  \bibinfo {pages} {2508} (\bibinfo {year} {1994})}\BibitemShut {NoStop}%
\bibitem [{\citenamefont {Huelga}\ \emph {et~al.}(1997)\citenamefont {Huelga},
  \citenamefont {Macchiavello}, \citenamefont {Pellizzari}, \citenamefont
  {Ekert}, \citenamefont {Plenio},\ and\ \citenamefont {Cirac}}]{Huelga1997}%
  \BibitemOpen
  \bibfield  {author} {\bibinfo {author} {\bibfnamefont {S.~F.}\ \bibnamefont
  {Huelga}}, \bibinfo {author} {\bibfnamefont {C.}~\bibnamefont
  {Macchiavello}}, \bibinfo {author} {\bibfnamefont {T.}~\bibnamefont
  {Pellizzari}}, \bibinfo {author} {\bibfnamefont {A.~K.}\ \bibnamefont
  {Ekert}}, \bibinfo {author} {\bibfnamefont {M.~B.}\ \bibnamefont {Plenio}}, \
  and\ \bibinfo {author} {\bibfnamefont {J.~I.}\ \bibnamefont {Cirac}},\ }\href
  {\doibase 10.1103/PhysRevLett.79.3865} {\bibfield  {journal} {\bibinfo
  {journal} {Phys. Rev. Lett.}\ }\textbf {\bibinfo {volume} {79}},\ \bibinfo
  {pages} {3865} (\bibinfo {year} {1997})}\BibitemShut {NoStop}%
\bibitem [{\citenamefont {Shor}(1995)}]{Shor1995}%
  \BibitemOpen
  \bibfield  {author} {\bibinfo {author} {\bibfnamefont {P.~W.}\ \bibnamefont
  {Shor}},\ }\href {\doibase 10.1103/PhysRevA.52.R2493} {\bibfield  {journal}
  {\bibinfo  {journal} {Phys. Rev. A}\ }\textbf {\bibinfo {volume} {52}},\
  \bibinfo {pages} {R2493} (\bibinfo {year} {1995})}\BibitemShut {NoStop}%
\bibitem [{\citenamefont {Palma}\ \emph {et~al.}(1996)\citenamefont {Palma},
  \citenamefont {Suominen},\ and\ \citenamefont {Ekert}}]{Palma1996}%
  \BibitemOpen
  \bibfield  {author} {\bibinfo {author} {\bibfnamefont {G.~M.}\ \bibnamefont
  {Palma}}, \bibinfo {author} {\bibfnamefont {K.-A.}\ \bibnamefont {Suominen}},
  \ and\ \bibinfo {author} {\bibfnamefont {A.~K.}\ \bibnamefont {Ekert}},\
  }\href {\doibase 10.1098/rspa.1996.0029} {\bibfield  {journal} {\bibinfo
  {journal} {Proc. Royal Soc. Lond. A}\ }\textbf {\bibinfo {volume} {452}},\
  \bibinfo {pages} {567} (\bibinfo {year} {1996})}\BibitemShut {NoStop}%
\bibitem [{\citenamefont {Joos}\ and\ \citenamefont {Zeh}(1985)}]{Joos1985}%
  \BibitemOpen
  \bibfield  {author} {\bibinfo {author} {\bibfnamefont {E.}~\bibnamefont
  {Joos}}\ and\ \bibinfo {author} {\bibfnamefont {H.~D.}\ \bibnamefont {Zeh}},\
  }\href {\doibase 10.1007/BF01725541} {\bibfield  {journal} {\bibinfo
  {journal} {Zeitschrift f{\"u}r Physik B Condensed Matter}\ }\textbf {\bibinfo
  {volume} {59}},\ \bibinfo {pages} {223} (\bibinfo {year} {1985})}\BibitemShut
  {NoStop}%
\bibitem [{\citenamefont {Zurek}(2003)}]{Zurek2003}%
  \BibitemOpen
  \bibfield  {author} {\bibinfo {author} {\bibfnamefont {W.~H.}\ \bibnamefont
  {Zurek}},\ }\href {\doibase 10.1103/RevModPhys.75.715} {\bibfield  {journal}
  {\bibinfo  {journal} {Rev. Mod. Phys.}\ }\textbf {\bibinfo {volume} {75}},\
  \bibinfo {pages} {715} (\bibinfo {year} {2003})}\BibitemShut {NoStop}%
\bibitem [{\citenamefont {Myatt}\ \emph {et~al.}(2000)\citenamefont {Myatt},
  \citenamefont {King}, \citenamefont {Turchette}, \citenamefont {Sackett},
  \citenamefont {Kielpinski}, \citenamefont {Itano}, \citenamefont {Monroe},\
  and\ \citenamefont {Wineland}}]{Myatt2000}%
  \BibitemOpen
  \bibfield  {author} {\bibinfo {author} {\bibfnamefont {C.~J.}\ \bibnamefont
  {Myatt}}, \bibinfo {author} {\bibfnamefont {B.~E.}\ \bibnamefont {King}},
  \bibinfo {author} {\bibfnamefont {Q.~A.}\ \bibnamefont {Turchette}}, \bibinfo
  {author} {\bibfnamefont {C.~A.}\ \bibnamefont {Sackett}}, \bibinfo {author}
  {\bibfnamefont {D.}~\bibnamefont {Kielpinski}}, \bibinfo {author}
  {\bibfnamefont {W.~M.}\ \bibnamefont {Itano}}, \bibinfo {author}
  {\bibfnamefont {C.}~\bibnamefont {Monroe}}, \ and\ \bibinfo {author}
  {\bibfnamefont {D.~J.}\ \bibnamefont {Wineland}},\ }\href {\doibase
  10.1038/35002001} {\bibfield  {journal} {\bibinfo  {journal} {Nature}\
  }\textbf {\bibinfo {volume} {403}},\ \bibinfo {pages} {269} (\bibinfo {year}
  {2000})}\BibitemShut {NoStop}%
\bibitem [{\citenamefont {Ratschbacher}\ \emph {et~al.}(2013)\citenamefont
  {Ratschbacher}, \citenamefont {Sias}, \citenamefont {Carcagni}, \citenamefont
  {Silver}, \citenamefont {Zipkes},\ and\ \citenamefont
  {K\"ohl}}]{Ratschbacher2013}%
  \BibitemOpen
  \bibfield  {author} {\bibinfo {author} {\bibfnamefont {L.}~\bibnamefont
  {Ratschbacher}}, \bibinfo {author} {\bibfnamefont {C.}~\bibnamefont {Sias}},
  \bibinfo {author} {\bibfnamefont {L.}~\bibnamefont {Carcagni}}, \bibinfo
  {author} {\bibfnamefont {J.~M.}\ \bibnamefont {Silver}}, \bibinfo {author}
  {\bibfnamefont {C.}~\bibnamefont {Zipkes}}, \ and\ \bibinfo {author}
  {\bibfnamefont {M.}~\bibnamefont {K\"ohl}},\ }\href {\doibase
  10.1103/PhysRevLett.110.160402} {\bibfield  {journal} {\bibinfo  {journal}
  {Phys. Rev. Lett.}\ }\textbf {\bibinfo {volume} {110}},\ \bibinfo {pages}
  {160402} (\bibinfo {year} {2013})}\BibitemShut {NoStop}%
\bibitem [{\citenamefont {Cetina}\ \emph {et~al.}(2016)\citenamefont {Cetina},
  \citenamefont {Jag}, \citenamefont {Lous}, \citenamefont {Fritsche},
  \citenamefont {Walraven}, \citenamefont {Grimm}, \citenamefont {Levinsen},
  \citenamefont {Parish}, \citenamefont {Schmidt}, \citenamefont {Knap},\ and\
  \citenamefont {Demler}}]{Cetina2016}%
  \BibitemOpen
  \bibfield  {author} {\bibinfo {author} {\bibfnamefont {M.}~\bibnamefont
  {Cetina}}, \bibinfo {author} {\bibfnamefont {M.}~\bibnamefont {Jag}},
  \bibinfo {author} {\bibfnamefont {R.~S.}\ \bibnamefont {Lous}}, \bibinfo
  {author} {\bibfnamefont {I.}~\bibnamefont {Fritsche}}, \bibinfo {author}
  {\bibfnamefont {J.~T.~M.}\ \bibnamefont {Walraven}}, \bibinfo {author}
  {\bibfnamefont {R.}~\bibnamefont {Grimm}}, \bibinfo {author} {\bibfnamefont
  {J.}~\bibnamefont {Levinsen}}, \bibinfo {author} {\bibfnamefont {M.~M.}\
  \bibnamefont {Parish}}, \bibinfo {author} {\bibfnamefont {R.}~\bibnamefont
  {Schmidt}}, \bibinfo {author} {\bibfnamefont {M.}~\bibnamefont {Knap}}, \
  and\ \bibinfo {author} {\bibfnamefont {E.}~\bibnamefont {Demler}},\ }\href
  {\doibase 10.1126/science.aaf5134} {\bibfield  {journal} {\bibinfo  {journal}
  {Science}\ }\textbf {\bibinfo {volume} {354}},\ \bibinfo {pages} {96}
  (\bibinfo {year} {2016})}\BibitemShut {NoStop}%
\bibitem [{\citenamefont {Liu}\ \emph {et~al.}(2018)\citenamefont {Liu},
  \citenamefont {Lyyra}, \citenamefont {Sun}, \citenamefont {Liu},
  \citenamefont {Li}, \citenamefont {Guo}, \citenamefont {Maniscalco},\ and\
  \citenamefont {Piilo}}]{Liu2018}%
  \BibitemOpen
  \bibfield  {author} {\bibinfo {author} {\bibfnamefont {Z.-D.}\ \bibnamefont
  {Liu}}, \bibinfo {author} {\bibfnamefont {H.}~\bibnamefont {Lyyra}}, \bibinfo
  {author} {\bibfnamefont {Y.-N.}\ \bibnamefont {Sun}}, \bibinfo {author}
  {\bibfnamefont {B.-H.}\ \bibnamefont {Liu}}, \bibinfo {author} {\bibfnamefont
  {C.-F.}\ \bibnamefont {Li}}, \bibinfo {author} {\bibfnamefont {G.-C.}\
  \bibnamefont {Guo}}, \bibinfo {author} {\bibfnamefont {S.}~\bibnamefont
  {Maniscalco}}, \ and\ \bibinfo {author} {\bibfnamefont {J.}~\bibnamefont
  {Piilo}},\ }\href {\doibase 10.1038/s41467-018-05817-x} {\bibfield  {journal}
  {\bibinfo  {journal} {Nature Commun.}\ }\textbf {\bibinfo {volume} {9}},\
  \bibinfo {pages} {3453} (\bibinfo {year} {2018})}\BibitemShut {NoStop}%
\bibitem [{\citenamefont {Unden}\ \emph {et~al.}(2019)\citenamefont {Unden},
  \citenamefont {Louzon}, \citenamefont {Zwolak}, \citenamefont {Zurek},\ and\
  \citenamefont {Jelezko}}]{Unden2019}%
  \BibitemOpen
  \bibfield  {author} {\bibinfo {author} {\bibfnamefont {T.~K.}\ \bibnamefont
  {Unden}}, \bibinfo {author} {\bibfnamefont {D.}~\bibnamefont {Louzon}},
  \bibinfo {author} {\bibfnamefont {M.}~\bibnamefont {Zwolak}}, \bibinfo
  {author} {\bibfnamefont {W.~H.}\ \bibnamefont {Zurek}}, \ and\ \bibinfo
  {author} {\bibfnamefont {F.}~\bibnamefont {Jelezko}},\ }\href {\doibase
  10.1103/PhysRevLett.123.140402} {\bibfield  {journal} {\bibinfo  {journal}
  {Phys. Rev. Lett.}\ }\textbf {\bibinfo {volume} {123}},\ \bibinfo {pages}
  {140402} (\bibinfo {year} {2019})}\BibitemShut {NoStop}%
\bibitem [{\citenamefont {Skou}\ \emph {et~al.}(2021)\citenamefont {Skou},
  \citenamefont {Skov}, \citenamefont {J{\o}rgensen}, \citenamefont {Nielsen},
  \citenamefont {Camacho-Guardian}, \citenamefont {Pohl}, \citenamefont
  {Bruun},\ and\ \citenamefont {Arlt}}]{Skou2021}%
  \BibitemOpen
  \bibfield  {author} {\bibinfo {author} {\bibfnamefont {M.~G.}\ \bibnamefont
  {Skou}}, \bibinfo {author} {\bibfnamefont {T.~G.}\ \bibnamefont {Skov}},
  \bibinfo {author} {\bibfnamefont {N.~B.}\ \bibnamefont {J{\o}rgensen}},
  \bibinfo {author} {\bibfnamefont {K.~K.}\ \bibnamefont {Nielsen}}, \bibinfo
  {author} {\bibfnamefont {A.}~\bibnamefont {Camacho-Guardian}}, \bibinfo
  {author} {\bibfnamefont {T.}~\bibnamefont {Pohl}}, \bibinfo {author}
  {\bibfnamefont {G.~M.}\ \bibnamefont {Bruun}}, \ and\ \bibinfo {author}
  {\bibfnamefont {J.~J.}\ \bibnamefont {Arlt}},\ }\href {\doibase
  10.1038/s41567-021-01184-5} {\bibfield  {journal} {\bibinfo  {journal}
  {Nature Physics}\ }\textbf {\bibinfo {volume} {17}},\ \bibinfo {pages} {731}
  (\bibinfo {year} {2021})}\BibitemShut {NoStop}%
\bibitem [{\citenamefont {Quan}\ \emph {et~al.}(2006)\citenamefont {Quan},
  \citenamefont {Song}, \citenamefont {Liu}, \citenamefont {Zanardi},\ and\
  \citenamefont {Sun}}]{Quan2006}%
  \BibitemOpen
  \bibfield  {author} {\bibinfo {author} {\bibfnamefont {H.~T.}\ \bibnamefont
  {Quan}}, \bibinfo {author} {\bibfnamefont {Z.}~\bibnamefont {Song}}, \bibinfo
  {author} {\bibfnamefont {X.~F.}\ \bibnamefont {Liu}}, \bibinfo {author}
  {\bibfnamefont {P.}~\bibnamefont {Zanardi}}, \ and\ \bibinfo {author}
  {\bibfnamefont {C.~P.}\ \bibnamefont {Sun}},\ }\href {\doibase
  10.1103/PhysRevLett.96.140604} {\bibfield  {journal} {\bibinfo  {journal}
  {Phys. Rev. Lett.}\ }\textbf {\bibinfo {volume} {96}},\ \bibinfo {pages}
  {140604} (\bibinfo {year} {2006})}\BibitemShut {NoStop}%
\bibitem [{\citenamefont {Bylander}\ \emph {et~al.}(2011)\citenamefont
  {Bylander}, \citenamefont {Gustavsson}, \citenamefont {Yan}, \citenamefont
  {Yoshihara}, \citenamefont {Harrabi}, \citenamefont {Fitch}, \citenamefont
  {Cory}, \citenamefont {Nakamura}, \citenamefont {Tsai},\ and\ \citenamefont
  {Oliver}}]{Bylander2011}%
  \BibitemOpen
  \bibfield  {author} {\bibinfo {author} {\bibfnamefont {J.}~\bibnamefont
  {Bylander}}, \bibinfo {author} {\bibfnamefont {S.}~\bibnamefont
  {Gustavsson}}, \bibinfo {author} {\bibfnamefont {F.}~\bibnamefont {Yan}},
  \bibinfo {author} {\bibfnamefont {F.}~\bibnamefont {Yoshihara}}, \bibinfo
  {author} {\bibfnamefont {K.}~\bibnamefont {Harrabi}}, \bibinfo {author}
  {\bibfnamefont {G.}~\bibnamefont {Fitch}}, \bibinfo {author} {\bibfnamefont
  {D.~G.}\ \bibnamefont {Cory}}, \bibinfo {author} {\bibfnamefont
  {Y.}~\bibnamefont {Nakamura}}, \bibinfo {author} {\bibfnamefont {J.-S.}\
  \bibnamefont {Tsai}}, \ and\ \bibinfo {author} {\bibfnamefont {W.~D.}\
  \bibnamefont {Oliver}},\ }\href {\doibase 10.1038/nphys1994} {\bibfield
  {journal} {\bibinfo  {journal} {Nature Physics}\ }\textbf {\bibinfo {volume}
  {7}},\ \bibinfo {pages} {565} (\bibinfo {year} {2011})}\BibitemShut {NoStop}%
\bibitem [{\citenamefont {Dorner}\ \emph {et~al.}(2013)\citenamefont {Dorner},
  \citenamefont {Clark}, \citenamefont {Heaney}, \citenamefont {Fazio},
  \citenamefont {Goold},\ and\ \citenamefont {Vedral}}]{Dorner2013}%
  \BibitemOpen
  \bibfield  {author} {\bibinfo {author} {\bibfnamefont {R.}~\bibnamefont
  {Dorner}}, \bibinfo {author} {\bibfnamefont {S.~R.}\ \bibnamefont {Clark}},
  \bibinfo {author} {\bibfnamefont {L.}~\bibnamefont {Heaney}}, \bibinfo
  {author} {\bibfnamefont {R.}~\bibnamefont {Fazio}}, \bibinfo {author}
  {\bibfnamefont {J.}~\bibnamefont {Goold}}, \ and\ \bibinfo {author}
  {\bibfnamefont {V.}~\bibnamefont {Vedral}},\ }\href {\doibase
  10.1103/PhysRevLett.110.230601} {\bibfield  {journal} {\bibinfo  {journal}
  {Phys. Rev. Lett.}\ }\textbf {\bibinfo {volume} {110}},\ \bibinfo {pages}
  {230601} (\bibinfo {year} {2013})}\BibitemShut {NoStop}%
\bibitem [{\citenamefont {Mazzola}\ \emph {et~al.}(2013)\citenamefont
  {Mazzola}, \citenamefont {De~Chiara},\ and\ \citenamefont
  {Paternostro}}]{Mazzola2013}%
  \BibitemOpen
  \bibfield  {author} {\bibinfo {author} {\bibfnamefont {L.}~\bibnamefont
  {Mazzola}}, \bibinfo {author} {\bibfnamefont {G.}~\bibnamefont {De~Chiara}},
  \ and\ \bibinfo {author} {\bibfnamefont {M.}~\bibnamefont {Paternostro}},\
  }\href {\doibase 10.1103/PhysRevLett.110.230602} {\bibfield  {journal}
  {\bibinfo  {journal} {Phys. Rev. Lett.}\ }\textbf {\bibinfo {volume} {110}},\
  \bibinfo {pages} {230602} (\bibinfo {year} {2013})}\BibitemShut {NoStop}%
\bibitem [{\citenamefont {Romach}\ \emph {et~al.}(2015)\citenamefont {Romach},
  \citenamefont {M\"uller}, \citenamefont {Unden}, \citenamefont {Rogers},
  \citenamefont {Isoda}, \citenamefont {Itoh}, \citenamefont {Markham},
  \citenamefont {Stacey}, \citenamefont {Meijer}, \citenamefont {Pezzagna},
  \citenamefont {Naydenov}, \citenamefont {McGuinness}, \citenamefont
  {Bar-Gill},\ and\ \citenamefont {Jelezko}}]{Romach2015}%
  \BibitemOpen
  \bibfield  {author} {\bibinfo {author} {\bibfnamefont {Y.}~\bibnamefont
  {Romach}}, \bibinfo {author} {\bibfnamefont {C.}~\bibnamefont {M\"uller}},
  \bibinfo {author} {\bibfnamefont {T.}~\bibnamefont {Unden}}, \bibinfo
  {author} {\bibfnamefont {L.~J.}\ \bibnamefont {Rogers}}, \bibinfo {author}
  {\bibfnamefont {T.}~\bibnamefont {Isoda}}, \bibinfo {author} {\bibfnamefont
  {K.~M.}\ \bibnamefont {Itoh}}, \bibinfo {author} {\bibfnamefont
  {M.}~\bibnamefont {Markham}}, \bibinfo {author} {\bibfnamefont
  {A.}~\bibnamefont {Stacey}}, \bibinfo {author} {\bibfnamefont
  {J.}~\bibnamefont {Meijer}}, \bibinfo {author} {\bibfnamefont
  {S.}~\bibnamefont {Pezzagna}}, \bibinfo {author} {\bibfnamefont
  {B.}~\bibnamefont {Naydenov}}, \bibinfo {author} {\bibfnamefont {L.~P.}\
  \bibnamefont {McGuinness}}, \bibinfo {author} {\bibfnamefont
  {N.}~\bibnamefont {Bar-Gill}}, \ and\ \bibinfo {author} {\bibfnamefont
  {F.}~\bibnamefont {Jelezko}},\ }\href {\doibase
  10.1103/PhysRevLett.114.017601} {\bibfield  {journal} {\bibinfo  {journal}
  {Phys. Rev. Lett.}\ }\textbf {\bibinfo {volume} {114}},\ \bibinfo {pages}
  {017601} (\bibinfo {year} {2015})}\BibitemShut {NoStop}%
\bibitem [{\citenamefont {Norris}\ \emph {et~al.}(2016)\citenamefont {Norris},
  \citenamefont {Paz-Silva},\ and\ \citenamefont {Viola}}]{Norris2016}%
  \BibitemOpen
  \bibfield  {author} {\bibinfo {author} {\bibfnamefont {L.~M.}\ \bibnamefont
  {Norris}}, \bibinfo {author} {\bibfnamefont {G.~A.}\ \bibnamefont
  {Paz-Silva}}, \ and\ \bibinfo {author} {\bibfnamefont {L.}~\bibnamefont
  {Viola}},\ }\href {\doibase 10.1103/PhysRevLett.116.150503} {\bibfield
  {journal} {\bibinfo  {journal} {Phys. Rev. Lett.}\ }\textbf {\bibinfo
  {volume} {116}},\ \bibinfo {pages} {150503} (\bibinfo {year}
  {2016})}\BibitemShut {NoStop}%
\bibitem [{\citenamefont {Sung}\ \emph {et~al.}(2019)\citenamefont {Sung},
  \citenamefont {Beaudoin}, \citenamefont {Norris}, \citenamefont {Yan},
  \citenamefont {Kim}, \citenamefont {Qiu}, \citenamefont {von Lüpke},
  \citenamefont {Yoder}, \citenamefont {Orlando}, \citenamefont {Gustavsson},
  \citenamefont {Viola},\ and\ \citenamefont {Oliver}}]{Sung2019}%
  \BibitemOpen
  \bibfield  {author} {\bibinfo {author} {\bibfnamefont {Y.}~\bibnamefont
  {Sung}}, \bibinfo {author} {\bibfnamefont {F.}~\bibnamefont {Beaudoin}},
  \bibinfo {author} {\bibfnamefont {L.~M.}\ \bibnamefont {Norris}}, \bibinfo
  {author} {\bibfnamefont {F.}~\bibnamefont {Yan}}, \bibinfo {author}
  {\bibfnamefont {D.~K.}\ \bibnamefont {Kim}}, \bibinfo {author} {\bibfnamefont
  {J.~Y.}\ \bibnamefont {Qiu}}, \bibinfo {author} {\bibfnamefont
  {U.}~\bibnamefont {von Lüpke}}, \bibinfo {author} {\bibfnamefont {J.~L.}\
  \bibnamefont {Yoder}}, \bibinfo {author} {\bibfnamefont {T.~P.}\ \bibnamefont
  {Orlando}}, \bibinfo {author} {\bibfnamefont {S.}~\bibnamefont {Gustavsson}},
  \bibinfo {author} {\bibfnamefont {L.}~\bibnamefont {Viola}}, \ and\ \bibinfo
  {author} {\bibfnamefont {W.~D.}\ \bibnamefont {Oliver}},\ }\href {\doibase
  10.1038/s41467-019-11699-4} {\bibfield  {journal} {\bibinfo  {journal}
  {Nature Commun.}\ }\textbf {\bibinfo {volume} {10}},\ \bibinfo {pages} {3715}
  (\bibinfo {year} {2019})}\BibitemShut {NoStop}%
\bibitem [{\citenamefont {Mitchison}\ \emph {et~al.}(2020)\citenamefont
  {Mitchison}, \citenamefont {Fogarty}, \citenamefont {Guarnieri},
  \citenamefont {Campbell}, \citenamefont {Busch},\ and\ \citenamefont
  {Goold}}]{Mitchison2020}%
  \BibitemOpen
  \bibfield  {author} {\bibinfo {author} {\bibfnamefont {M.~T.}\ \bibnamefont
  {Mitchison}}, \bibinfo {author} {\bibfnamefont {T.}~\bibnamefont {Fogarty}},
  \bibinfo {author} {\bibfnamefont {G.}~\bibnamefont {Guarnieri}}, \bibinfo
  {author} {\bibfnamefont {S.}~\bibnamefont {Campbell}}, \bibinfo {author}
  {\bibfnamefont {T.}~\bibnamefont {Busch}}, \ and\ \bibinfo {author}
  {\bibfnamefont {J.}~\bibnamefont {Goold}},\ }\href {\doibase
  10.1103/PhysRevLett.125.080402} {\bibfield  {journal} {\bibinfo  {journal}
  {Phys. Rev. Lett.}\ }\textbf {\bibinfo {volume} {125}},\ \bibinfo {pages}
  {080402} (\bibinfo {year} {2020})}\BibitemShut {NoStop}%
\bibitem [{\citenamefont {Mitchison}\ \emph {et~al.}(2022)\citenamefont
  {Mitchison}, \citenamefont {Purkayastha}, \citenamefont {Brenes},
  \citenamefont {Silva},\ and\ \citenamefont {Goold}}]{Mitchison2021}%
  \BibitemOpen
  \bibfield  {author} {\bibinfo {author} {\bibfnamefont {M.~T.}\ \bibnamefont
  {Mitchison}}, \bibinfo {author} {\bibfnamefont {A.}~\bibnamefont
  {Purkayastha}}, \bibinfo {author} {\bibfnamefont {M.}~\bibnamefont {Brenes}},
  \bibinfo {author} {\bibfnamefont {A.}~\bibnamefont {Silva}}, \ and\ \bibinfo
  {author} {\bibfnamefont {J.}~\bibnamefont {Goold}},\ }\href {\doibase
  10.1103/PhysRevA.105.L030201} {\bibfield  {journal} {\bibinfo  {journal}
  {Phys. Rev. A}\ }\textbf {\bibinfo {volume} {105}},\ \bibinfo {pages}
  {L030201} (\bibinfo {year} {2022})}\BibitemShut {NoStop}%
\bibitem [{\citenamefont {Adam}\ \emph {et~al.}(2022)\citenamefont {Adam},
  \citenamefont {Bouton}, \citenamefont {Nettersheim}, \citenamefont
  {Burgardt},\ and\ \citenamefont {Widera}}]{Adam2022}%
  \BibitemOpen
  \bibfield  {author} {\bibinfo {author} {\bibfnamefont {D.}~\bibnamefont
  {Adam}}, \bibinfo {author} {\bibfnamefont {Q.}~\bibnamefont {Bouton}},
  \bibinfo {author} {\bibfnamefont {J.}~\bibnamefont {Nettersheim}}, \bibinfo
  {author} {\bibfnamefont {S.}~\bibnamefont {Burgardt}}, \ and\ \bibinfo
  {author} {\bibfnamefont {A.}~\bibnamefont {Widera}},\ }\href {\doibase
  10.1103/PhysRevLett.129.120404} {\bibfield  {journal} {\bibinfo  {journal}
  {Phys. Rev. Lett.}\ }\textbf {\bibinfo {volume} {129}},\ \bibinfo {pages}
  {120404} (\bibinfo {year} {2022})}\BibitemShut {NoStop}%
\bibitem [{\citenamefont {Spohn}(1978)}]{Spohn1978a}%
  \BibitemOpen
  \bibfield  {author} {\bibinfo {author} {\bibfnamefont {H.}~\bibnamefont
  {Spohn}},\ }\href {\doibase 10.1063/1.523789} {\bibfield  {journal} {\bibinfo
   {journal} {Journal of Mathematical Physics}\ }\textbf {\bibinfo {volume}
  {19}},\ \bibinfo {pages} {1227} (\bibinfo {year} {1978})}\BibitemShut
  {NoStop}%
\bibitem [{\citenamefont {Esposito}\ \emph {et~al.}(2010)\citenamefont
  {Esposito}, \citenamefont {Lindenberg},\ and\ \citenamefont {den
  Broeck}}]{Esposito2010}%
  \BibitemOpen
  \bibfield  {author} {\bibinfo {author} {\bibfnamefont {M.}~\bibnamefont
  {Esposito}}, \bibinfo {author} {\bibfnamefont {K.}~\bibnamefont
  {Lindenberg}}, \ and\ \bibinfo {author} {\bibfnamefont {C.~V.}\ \bibnamefont
  {den Broeck}},\ }\href {\doibase 10.1088/1367-2630/12/1/013013} {\bibfield
  {journal} {\bibinfo  {journal} {New Journal of Physics}\ }\textbf {\bibinfo
  {volume} {12}},\ \bibinfo {pages} {013013} (\bibinfo {year}
  {2010})}\BibitemShut {NoStop}%
\bibitem [{\citenamefont {Silva}(2008)}]{Silva2008}%
  \BibitemOpen
  \bibfield  {author} {\bibinfo {author} {\bibfnamefont {A.}~\bibnamefont
  {Silva}},\ }\href {\doibase 10.1103/PhysRevLett.101.120603} {\bibfield
  {journal} {\bibinfo  {journal} {Phys. Rev. Lett.}\ }\textbf {\bibinfo
  {volume} {101}},\ \bibinfo {pages} {120603} (\bibinfo {year}
  {2008})}\BibitemShut {NoStop}%
\bibitem [{\citenamefont {Smacchia}\ and\ \citenamefont
  {Silva}(2012)}]{Smacchia2012}%
  \BibitemOpen
  \bibfield  {author} {\bibinfo {author} {\bibfnamefont {P.}~\bibnamefont
  {Smacchia}}\ and\ \bibinfo {author} {\bibfnamefont {A.}~\bibnamefont
  {Silva}},\ }\href {\doibase 10.1103/PhysRevLett.109.037202} {\bibfield
  {journal} {\bibinfo  {journal} {Phys. Rev. Lett.}\ }\textbf {\bibinfo
  {volume} {109}},\ \bibinfo {pages} {037202} (\bibinfo {year}
  {2012})}\BibitemShut {NoStop}%
\bibitem [{\citenamefont {Sindona}\ \emph {et~al.}(2014)\citenamefont
  {Sindona}, \citenamefont {Goold}, \citenamefont {Gullo},\ and\ \citenamefont
  {Plastina}}]{Sindona2014}%
  \BibitemOpen
  \bibfield  {author} {\bibinfo {author} {\bibfnamefont {A.}~\bibnamefont
  {Sindona}}, \bibinfo {author} {\bibfnamefont {J.}~\bibnamefont {Goold}},
  \bibinfo {author} {\bibfnamefont {N.~L.}\ \bibnamefont {Gullo}}, \ and\
  \bibinfo {author} {\bibfnamefont {F.}~\bibnamefont {Plastina}},\ }\href
  {\doibase 10.1088/1367-2630/16/4/045013} {\bibfield  {journal} {\bibinfo
  {journal} {New J. Phys.}\ }\textbf {\bibinfo {volume} {16}},\ \bibinfo
  {pages} {045013} (\bibinfo {year} {2014})}\BibitemShut {NoStop}%
\bibitem [{Note1()}]{Note1}%
  \BibitemOpen
  \bibinfo {note} {This identifies the system energy eigenstates $\protect
  \{{\left \delimiter 69640972 n\right \delimiter "526930B }\protect \}$ as the
  pointer states selected by the environment.}\BibitemShut {Stop}%
\bibitem [{\citenamefont {Talkner}\ \emph {et~al.}(2007)\citenamefont
  {Talkner}, \citenamefont {Lutz},\ and\ \citenamefont
  {H\"anggi}}]{Talkner2007}%
  \BibitemOpen
  \bibfield  {author} {\bibinfo {author} {\bibfnamefont {P.}~\bibnamefont
  {Talkner}}, \bibinfo {author} {\bibfnamefont {E.}~\bibnamefont {Lutz}}, \
  and\ \bibinfo {author} {\bibfnamefont {P.}~\bibnamefont {H\"anggi}},\ }\href
  {\doibase 10.1103/PhysRevE.75.050102} {\bibfield  {journal} {\bibinfo
  {journal} {Phys. Rev. E}\ }\textbf {\bibinfo {volume} {75}},\ \bibinfo
  {pages} {050102} (\bibinfo {year} {2007})}\BibitemShut {NoStop}%
\bibitem [{\citenamefont {Esposito}\ \emph {et~al.}(2009)\citenamefont
  {Esposito}, \citenamefont {Harbola},\ and\ \citenamefont
  {Mukamel}}]{Esposito2009}%
  \BibitemOpen
  \bibfield  {author} {\bibinfo {author} {\bibfnamefont {M.}~\bibnamefont
  {Esposito}}, \bibinfo {author} {\bibfnamefont {U.}~\bibnamefont {Harbola}}, \
  and\ \bibinfo {author} {\bibfnamefont {S.}~\bibnamefont {Mukamel}},\ }\href
  {\doibase 10.1103/RevModPhys.81.1665} {\bibfield  {journal} {\bibinfo
  {journal} {Rev. Mod. Phys.}\ }\textbf {\bibinfo {volume} {81}},\ \bibinfo
  {pages} {1665} (\bibinfo {year} {2009})}\BibitemShut {NoStop}%
\bibitem [{\citenamefont {Campisi}\ \emph {et~al.}(2011)\citenamefont
  {Campisi}, \citenamefont {H\"anggi},\ and\ \citenamefont
  {Talkner}}]{Campisi2011}%
  \BibitemOpen
  \bibfield  {author} {\bibinfo {author} {\bibfnamefont {M.}~\bibnamefont
  {Campisi}}, \bibinfo {author} {\bibfnamefont {P.}~\bibnamefont {H\"anggi}}, \
  and\ \bibinfo {author} {\bibfnamefont {P.}~\bibnamefont {Talkner}},\ }\href
  {\doibase 10.1103/RevModPhys.83.771} {\bibfield  {journal} {\bibinfo
  {journal} {Rev. Mod. Phys.}\ }\textbf {\bibinfo {volume} {83}},\ \bibinfo
  {pages} {771} (\bibinfo {year} {2011})}\BibitemShut {NoStop}%
\bibitem [{\citenamefont {Talkner}\ and\ \citenamefont
  {H\"anggi}(2020)}]{Talkner2020}%
  \BibitemOpen
  \bibfield  {author} {\bibinfo {author} {\bibfnamefont {P.}~\bibnamefont
  {Talkner}}\ and\ \bibinfo {author} {\bibfnamefont {P.}~\bibnamefont
  {H\"anggi}},\ }\href {\doibase 10.1103/RevModPhys.92.041002} {\bibfield
  {journal} {\bibinfo  {journal} {Rev. Mod. Phys.}\ }\textbf {\bibinfo {volume}
  {92}},\ \bibinfo {pages} {041002} (\bibinfo {year} {2020})}\BibitemShut
  {NoStop}%
\bibitem [{\citenamefont {Landi}\ and\ \citenamefont
  {Paternostro}(2021)}]{Landi2021}%
  \BibitemOpen
  \bibfield  {author} {\bibinfo {author} {\bibfnamefont {G.~T.}\ \bibnamefont
  {Landi}}\ and\ \bibinfo {author} {\bibfnamefont {M.}~\bibnamefont
  {Paternostro}},\ }\href {\doibase 10.1103/RevModPhys.93.035008} {\bibfield
  {journal} {\bibinfo  {journal} {Rev. Mod. Phys.}\ }\textbf {\bibinfo {volume}
  {93}},\ \bibinfo {pages} {035008} (\bibinfo {year} {2021})}\BibitemShut
  {NoStop}%
\bibitem [{\citenamefont {Goold}\ \emph {et~al.}(2014)\citenamefont {Goold},
  \citenamefont {Poschinger},\ and\ \citenamefont {Modi}}]{Goold2014}%
  \BibitemOpen
  \bibfield  {author} {\bibinfo {author} {\bibfnamefont {J.}~\bibnamefont
  {Goold}}, \bibinfo {author} {\bibfnamefont {U.}~\bibnamefont {Poschinger}}, \
  and\ \bibinfo {author} {\bibfnamefont {K.}~\bibnamefont {Modi}},\ }\href
  {\doibase 10.1103/PhysRevE.90.020101} {\bibfield  {journal} {\bibinfo
  {journal} {Phys. Rev. E}\ }\textbf {\bibinfo {volume} {90}},\ \bibinfo
  {pages} {020101} (\bibinfo {year} {2014})}\BibitemShut {NoStop}%
\bibitem [{\citenamefont {Funo}\ and\ \citenamefont {Quan}(2018)}]{Funo2018}%
  \BibitemOpen
  \bibfield  {author} {\bibinfo {author} {\bibfnamefont {K.}~\bibnamefont
  {Funo}}\ and\ \bibinfo {author} {\bibfnamefont {H.~T.}\ \bibnamefont
  {Quan}},\ }\href {\doibase 10.1103/PhysRevLett.121.040602} {\bibfield
  {journal} {\bibinfo  {journal} {Phys. Rev. Lett.}\ }\textbf {\bibinfo
  {volume} {121}},\ \bibinfo {pages} {040602} (\bibinfo {year}
  {2018})}\BibitemShut {NoStop}%
\bibitem [{\citenamefont {Popovic}\ \emph {et~al.}(2021)\citenamefont
  {Popovic}, \citenamefont {Mitchison}, \citenamefont {Strathearn},
  \citenamefont {Lovett}, \citenamefont {Goold},\ and\ \citenamefont
  {Eastham}}]{Popovic2021}%
  \BibitemOpen
  \bibfield  {author} {\bibinfo {author} {\bibfnamefont {M.}~\bibnamefont
  {Popovic}}, \bibinfo {author} {\bibfnamefont {M.~T.}\ \bibnamefont
  {Mitchison}}, \bibinfo {author} {\bibfnamefont {A.}~\bibnamefont
  {Strathearn}}, \bibinfo {author} {\bibfnamefont {B.~W.}\ \bibnamefont
  {Lovett}}, \bibinfo {author} {\bibfnamefont {J.}~\bibnamefont {Goold}}, \
  and\ \bibinfo {author} {\bibfnamefont {P.~R.}\ \bibnamefont {Eastham}},\
  }\href {\doibase 10.1103/PRXQuantum.2.020338} {\bibfield  {journal} {\bibinfo
   {journal} {PRX Quantum}\ }\textbf {\bibinfo {volume} {2}},\ \bibinfo {pages}
  {020338} (\bibinfo {year} {2021})}\BibitemShut {NoStop}%
\bibitem [{\citenamefont {Jarzynski}(1997)}]{Jarzynski1997}%
  \BibitemOpen
  \bibfield  {author} {\bibinfo {author} {\bibfnamefont {C.}~\bibnamefont
  {Jarzynski}},\ }\href {\doibase 10.1103/PhysRevLett.78.2690} {\bibfield
  {journal} {\bibinfo  {journal} {Phys. Rev. Lett.}\ }\textbf {\bibinfo
  {volume} {78}},\ \bibinfo {pages} {2690} (\bibinfo {year}
  {1997})}\BibitemShut {NoStop}%
\bibitem [{\citenamefont {Goold}\ \emph {et~al.}(2015)\citenamefont {Goold},
  \citenamefont {Paternostro},\ and\ \citenamefont {Modi}}]{Goold2015}%
  \BibitemOpen
  \bibfield  {author} {\bibinfo {author} {\bibfnamefont {J.}~\bibnamefont
  {Goold}}, \bibinfo {author} {\bibfnamefont {M.}~\bibnamefont {Paternostro}},
  \ and\ \bibinfo {author} {\bibfnamefont {K.}~\bibnamefont {Modi}},\ }\href
  {\doibase 10.1103/PhysRevLett.114.060602} {\bibfield  {journal} {\bibinfo
  {journal} {Phys. Rev. Lett.}\ }\textbf {\bibinfo {volume} {114}},\ \bibinfo
  {pages} {060602} (\bibinfo {year} {2015})}\BibitemShut {NoStop}%
\bibitem [{\citenamefont {Sagawa}(2012)}]{Sagawa2012}%
  \BibitemOpen
  \bibfield  {author} {\bibinfo {author} {\bibfnamefont {T.}~\bibnamefont
  {Sagawa}},\ }in\ \href {\doibase 10.1142/9789814425193_0003} {\emph {\bibinfo
  {booktitle} {Kinki University Series on Quantum Computing}}}\ (\bibinfo
  {publisher} {World Scientific},\ \bibinfo {year} {2012})\ pp.\ \bibinfo
  {pages} {125--190}\BibitemShut {NoStop}%
\bibitem [{\citenamefont {Francica}\ \emph {et~al.}(2019)\citenamefont
  {Francica}, \citenamefont {Goold},\ and\ \citenamefont
  {Plastina}}]{Francica2019}%
  \BibitemOpen
  \bibfield  {author} {\bibinfo {author} {\bibfnamefont {G.}~\bibnamefont
  {Francica}}, \bibinfo {author} {\bibfnamefont {J.}~\bibnamefont {Goold}}, \
  and\ \bibinfo {author} {\bibfnamefont {F.}~\bibnamefont {Plastina}},\ }\href
  {\doibase 10.1103/PhysRevE.99.042105} {\bibfield  {journal} {\bibinfo
  {journal} {Phys. Rev. E}\ }\textbf {\bibinfo {volume} {99}},\ \bibinfo
  {pages} {042105} (\bibinfo {year} {2019})}\BibitemShut {NoStop}%
\bibitem [{\citenamefont {Santos}\ \emph {et~al.}(2019)\citenamefont {Santos},
  \citenamefont {C{\'{e}}leri}, \citenamefont {Landi},\ and\ \citenamefont
  {Paternostro}}]{Santos2019}%
  \BibitemOpen
  \bibfield  {author} {\bibinfo {author} {\bibfnamefont {J.~P.}\ \bibnamefont
  {Santos}}, \bibinfo {author} {\bibfnamefont {L.~C.}\ \bibnamefont
  {C{\'{e}}leri}}, \bibinfo {author} {\bibfnamefont {G.~T.}\ \bibnamefont
  {Landi}}, \ and\ \bibinfo {author} {\bibfnamefont {M.}~\bibnamefont
  {Paternostro}},\ }\href {\doibase 10.1038/s41534-019-0138-y} {\bibfield
  {journal} {\bibinfo  {journal} {npj Quantum Information}\ }\textbf {\bibinfo
  {volume} {5}},\ \bibinfo {pages} {23} (\bibinfo {year} {2019})}\BibitemShut
  {NoStop}%
\bibitem [{\citenamefont {Francica}\ \emph {et~al.}(2020)\citenamefont
  {Francica}, \citenamefont {Binder}, \citenamefont {Guarnieri}, \citenamefont
  {Mitchison}, \citenamefont {Goold},\ and\ \citenamefont
  {Plastina}}]{Francica2020}%
  \BibitemOpen
  \bibfield  {author} {\bibinfo {author} {\bibfnamefont {G.}~\bibnamefont
  {Francica}}, \bibinfo {author} {\bibfnamefont {F.~C.}\ \bibnamefont
  {Binder}}, \bibinfo {author} {\bibfnamefont {G.}~\bibnamefont {Guarnieri}},
  \bibinfo {author} {\bibfnamefont {M.~T.}\ \bibnamefont {Mitchison}}, \bibinfo
  {author} {\bibfnamefont {J.}~\bibnamefont {Goold}}, \ and\ \bibinfo {author}
  {\bibfnamefont {F.}~\bibnamefont {Plastina}},\ }\href {\doibase
  10.1103/PhysRevLett.125.180603} {\bibfield  {journal} {\bibinfo  {journal}
  {Phys. Rev. Lett.}\ }\textbf {\bibinfo {volume} {125}},\ \bibinfo {pages}
  {180603} (\bibinfo {year} {2020})}\BibitemShut {NoStop}%
\bibitem [{\citenamefont {Clerk}\ \emph {et~al.}(2010)\citenamefont {Clerk},
  \citenamefont {Devoret}, \citenamefont {Girvin}, \citenamefont {Marquardt},\
  and\ \citenamefont {Schoelkopf}}]{Clerk2010}%
  \BibitemOpen
  \bibfield  {author} {\bibinfo {author} {\bibfnamefont {A.~A.}\ \bibnamefont
  {Clerk}}, \bibinfo {author} {\bibfnamefont {M.~H.}\ \bibnamefont {Devoret}},
  \bibinfo {author} {\bibfnamefont {S.~M.}\ \bibnamefont {Girvin}}, \bibinfo
  {author} {\bibfnamefont {F.}~\bibnamefont {Marquardt}}, \ and\ \bibinfo
  {author} {\bibfnamefont {R.~J.}\ \bibnamefont {Schoelkopf}},\ }\href
  {\doibase 10.1103/RevModPhys.82.1155} {\bibfield  {journal} {\bibinfo
  {journal} {Rev. Mod. Phys.}\ }\textbf {\bibinfo {volume} {82}},\ \bibinfo
  {pages} {1155} (\bibinfo {year} {2010})}\BibitemShut {NoStop}%
\bibitem [{\citenamefont {\'Alvarez}\ and\ \citenamefont
  {Suter}(2011)}]{Alvarez2011}%
  \BibitemOpen
  \bibfield  {author} {\bibinfo {author} {\bibfnamefont {G.~A.}\ \bibnamefont
  {\'Alvarez}}\ and\ \bibinfo {author} {\bibfnamefont {D.}~\bibnamefont
  {Suter}},\ }\href {\doibase 10.1103/PhysRevLett.107.230501} {\bibfield
  {journal} {\bibinfo  {journal} {Phys. Rev. Lett.}\ }\textbf {\bibinfo
  {volume} {107}},\ \bibinfo {pages} {230501} (\bibinfo {year}
  {2011})}\BibitemShut {NoStop}%
\bibitem [{\citenamefont {LeBlanc}\ and\ \citenamefont
  {Thywissen}(2007)}]{LeBlanc2007}%
  \BibitemOpen
  \bibfield  {author} {\bibinfo {author} {\bibfnamefont {L.~J.}\ \bibnamefont
  {LeBlanc}}\ and\ \bibinfo {author} {\bibfnamefont {J.~H.}\ \bibnamefont
  {Thywissen}},\ }\href {\doibase 10.1103/PhysRevA.75.053612} {\bibfield
  {journal} {\bibinfo  {journal} {Phys. Rev. A}\ }\textbf {\bibinfo {volume}
  {75}},\ \bibinfo {pages} {053612} (\bibinfo {year} {2007})}\BibitemShut
  {NoStop}%
\bibitem [{\citenamefont {Chin}\ \emph {et~al.}(2010)\citenamefont {Chin},
  \citenamefont {Grimm}, \citenamefont {Julienne},\ and\ \citenamefont
  {Tiesinga}}]{Chin2010}%
  \BibitemOpen
  \bibfield  {author} {\bibinfo {author} {\bibfnamefont {C.}~\bibnamefont
  {Chin}}, \bibinfo {author} {\bibfnamefont {R.}~\bibnamefont {Grimm}},
  \bibinfo {author} {\bibfnamefont {P.}~\bibnamefont {Julienne}}, \ and\
  \bibinfo {author} {\bibfnamefont {E.}~\bibnamefont {Tiesinga}},\ }\href
  {\doibase 10.1103/RevModPhys.82.1225} {\bibfield  {journal} {\bibinfo
  {journal} {Rev. Mod. Phys.}\ }\textbf {\bibinfo {volume} {82}},\ \bibinfo
  {pages} {1225} (\bibinfo {year} {2010})}\BibitemShut {NoStop}%
\bibitem [{\citenamefont {Goold}\ \emph {et~al.}(2011)\citenamefont {Goold},
  \citenamefont {Fogarty}, \citenamefont {Lo~Gullo}, \citenamefont
  {Paternostro},\ and\ \citenamefont {Busch}}]{Goold2011}%
  \BibitemOpen
  \bibfield  {author} {\bibinfo {author} {\bibfnamefont {J.}~\bibnamefont
  {Goold}}, \bibinfo {author} {\bibfnamefont {T.}~\bibnamefont {Fogarty}},
  \bibinfo {author} {\bibfnamefont {N.}~\bibnamefont {Lo~Gullo}}, \bibinfo
  {author} {\bibfnamefont {M.}~\bibnamefont {Paternostro}}, \ and\ \bibinfo
  {author} {\bibfnamefont {T.}~\bibnamefont {Busch}},\ }\href {\doibase
  10.1103/PhysRevA.84.063632} {\bibfield  {journal} {\bibinfo  {journal} {Phys.
  Rev. A}\ }\textbf {\bibinfo {volume} {84}},\ \bibinfo {pages} {063632}
  (\bibinfo {year} {2011})}\BibitemShut {NoStop}%
\bibitem [{\citenamefont {Knap}\ \emph {et~al.}(2012)\citenamefont {Knap},
  \citenamefont {Shashi}, \citenamefont {Nishida}, \citenamefont {Imambekov},
  \citenamefont {Abanin},\ and\ \citenamefont {Demler}}]{Knap2012}%
  \BibitemOpen
  \bibfield  {author} {\bibinfo {author} {\bibfnamefont {M.}~\bibnamefont
  {Knap}}, \bibinfo {author} {\bibfnamefont {A.}~\bibnamefont {Shashi}},
  \bibinfo {author} {\bibfnamefont {Y.}~\bibnamefont {Nishida}}, \bibinfo
  {author} {\bibfnamefont {A.}~\bibnamefont {Imambekov}}, \bibinfo {author}
  {\bibfnamefont {D.~A.}\ \bibnamefont {Abanin}}, \ and\ \bibinfo {author}
  {\bibfnamefont {E.}~\bibnamefont {Demler}},\ }\href {\doibase
  10.1103/PhysRevX.2.041020} {\bibfield  {journal} {\bibinfo  {journal} {Phys.
  Rev. X}\ }\textbf {\bibinfo {volume} {2}},\ \bibinfo {pages} {041020}
  (\bibinfo {year} {2012})}\BibitemShut {NoStop}%
\bibitem [{\citenamefont {Klich}(2003)}]{Klich2003}%
  \BibitemOpen
  \bibfield  {author} {\bibinfo {author} {\bibfnamefont {I.}~\bibnamefont
  {Klich}},\ }in\ \href {\doibase 10.1007/978-94-010-0089-5_19} {\emph
  {\bibinfo {booktitle} {Quantum Noise in Mesoscopic Physics}}}\ (\bibinfo
  {publisher} {Springer Netherlands},\ \bibinfo {year} {2003})\ pp.\ \bibinfo
  {pages} {397--402}\BibitemShut {NoStop}%
\bibitem [{\citenamefont {Abanin}\ and\ \citenamefont
  {Levitov}(2005)}]{Levitov2005}%
  \BibitemOpen
  \bibfield  {author} {\bibinfo {author} {\bibfnamefont {D.~A.}\ \bibnamefont
  {Abanin}}\ and\ \bibinfo {author} {\bibfnamefont {L.~S.}\ \bibnamefont
  {Levitov}},\ }\href {\doibase 10.1103/PhysRevLett.94.186803} {\bibfield
  {journal} {\bibinfo  {journal} {Phys. Rev. Lett.}\ }\textbf {\bibinfo
  {volume} {94}},\ \bibinfo {pages} {186803} (\bibinfo {year}
  {2005})}\BibitemShut {NoStop}%
\bibitem [{\citenamefont {Schmidt}\ \emph {et~al.}(2018)\citenamefont
  {Schmidt}, \citenamefont {Knap}, \citenamefont {Ivanov}, \citenamefont {You},
  \citenamefont {Cetina},\ and\ \citenamefont {Demler}}]{Schmidt2018}%
  \BibitemOpen
  \bibfield  {author} {\bibinfo {author} {\bibfnamefont {R.}~\bibnamefont
  {Schmidt}}, \bibinfo {author} {\bibfnamefont {M.}~\bibnamefont {Knap}},
  \bibinfo {author} {\bibfnamefont {D.~A.}\ \bibnamefont {Ivanov}}, \bibinfo
  {author} {\bibfnamefont {J.-S.}\ \bibnamefont {You}}, \bibinfo {author}
  {\bibfnamefont {M.}~\bibnamefont {Cetina}}, \ and\ \bibinfo {author}
  {\bibfnamefont {E.}~\bibnamefont {Demler}},\ }\href {\doibase
  10.1088/1361-6633/aa9593} {\bibfield  {journal} {\bibinfo  {journal} {Rep.
  Prog. Phys.}\ }\textbf {\bibinfo {volume} {81}},\ \bibinfo {pages} {024401}
  (\bibinfo {year} {2018})}\BibitemShut {NoStop}%
\bibitem [{\citenamefont {Mahan}(1967)}]{Mahan1967}%
  \BibitemOpen
  \bibfield  {author} {\bibinfo {author} {\bibfnamefont {G.~D.}\ \bibnamefont
  {Mahan}},\ }\href {\doibase 10.1103/PhysRev.163.612} {\bibfield  {journal}
  {\bibinfo  {journal} {Phys. Rev.}\ }\textbf {\bibinfo {volume} {163}},\
  \bibinfo {pages} {612} (\bibinfo {year} {1967})}\BibitemShut {NoStop}%
\bibitem [{\citenamefont {Nozi\`eres}\ and\ \citenamefont
  {De~Dominicis}(1969)}]{Nozieres1969}%
  \BibitemOpen
  \bibfield  {author} {\bibinfo {author} {\bibfnamefont {P.}~\bibnamefont
  {Nozi\`eres}}\ and\ \bibinfo {author} {\bibfnamefont {C.~T.}\ \bibnamefont
  {De~Dominicis}},\ }\href {\doibase 10.1103/PhysRev.178.1097} {\bibfield
  {journal} {\bibinfo  {journal} {Phys. Rev.}\ }\textbf {\bibinfo {volume}
  {178}},\ \bibinfo {pages} {1097} (\bibinfo {year} {1969})}\BibitemShut
  {NoStop}%
\bibitem [{\citenamefont {Anderson}(1967)}]{Anderson1967}%
  \BibitemOpen
  \bibfield  {author} {\bibinfo {author} {\bibfnamefont {P.~W.}\ \bibnamefont
  {Anderson}},\ }\href {\doibase 10.1103/PhysRevLett.18.1049} {\bibfield
  {journal} {\bibinfo  {journal} {Phys. Rev. Lett.}\ }\textbf {\bibinfo
  {volume} {18}},\ \bibinfo {pages} {1049} (\bibinfo {year}
  {1967})}\BibitemShut {NoStop}%
\bibitem [{\citenamefont {Breuer}\ \emph {et~al.}(2016)\citenamefont {Breuer},
  \citenamefont {Laine}, \citenamefont {Piilo},\ and\ \citenamefont
  {Vacchini}}]{Breuer2016}%
  \BibitemOpen
  \bibfield  {author} {\bibinfo {author} {\bibfnamefont {H.-P.}\ \bibnamefont
  {Breuer}}, \bibinfo {author} {\bibfnamefont {E.-M.}\ \bibnamefont {Laine}},
  \bibinfo {author} {\bibfnamefont {J.}~\bibnamefont {Piilo}}, \ and\ \bibinfo
  {author} {\bibfnamefont {B.}~\bibnamefont {Vacchini}},\ }\href {\doibase
  10.1103/RevModPhys.88.021002} {\bibfield  {journal} {\bibinfo  {journal}
  {Rev. Mod. Phys.}\ }\textbf {\bibinfo {volume} {88}},\ \bibinfo {pages}
  {021002} (\bibinfo {year} {2016})}\BibitemShut {NoStop}%
\bibitem [{\citenamefont {Bernardo}(2020)}]{deLima2020}%
  \BibitemOpen
  \bibfield  {author} {\bibinfo {author} {\bibfnamefont {B.~d.~L.}\
  \bibnamefont {Bernardo}},\ }\href {\doibase 10.1103/PhysRevE.102.062152}
  {\bibfield  {journal} {\bibinfo  {journal} {Phys. Rev. E}\ }\textbf {\bibinfo
  {volume} {102}},\ \bibinfo {pages} {062152} (\bibinfo {year}
  {2020})}\BibitemShut {NoStop}%
\bibitem [{\citenamefont {Alipour}\ \emph {et~al.}(2022)\citenamefont
  {Alipour}, \citenamefont {Rezakhani}, \citenamefont {Chenu}, \citenamefont
  {del Campo},\ and\ \citenamefont {Ala-Nissila}}]{Alipour2022}%
  \BibitemOpen
  \bibfield  {author} {\bibinfo {author} {\bibfnamefont {S.}~\bibnamefont
  {Alipour}}, \bibinfo {author} {\bibfnamefont {A.~T.}\ \bibnamefont
  {Rezakhani}}, \bibinfo {author} {\bibfnamefont {A.}~\bibnamefont {Chenu}},
  \bibinfo {author} {\bibfnamefont {A.}~\bibnamefont {del Campo}}, \ and\
  \bibinfo {author} {\bibfnamefont {T.}~\bibnamefont {Ala-Nissila}},\ }\href
  {\doibase 10.1103/PhysRevA.105.L040201} {\bibfield  {journal} {\bibinfo
  {journal} {Phys. Rev. A}\ }\textbf {\bibinfo {volume} {105}},\ \bibinfo
  {pages} {L040201} (\bibinfo {year} {2022})}\BibitemShut {NoStop}%
\bibitem [{\citenamefont {Ahmadi}\ \emph {et~al.}(2023)\citenamefont {Ahmadi},
  \citenamefont {Salimi},\ and\ \citenamefont {Khorashad}}]{Ahmadi2023}%
  \BibitemOpen
  \bibfield  {author} {\bibinfo {author} {\bibfnamefont {B.}~\bibnamefont
  {Ahmadi}}, \bibinfo {author} {\bibfnamefont {S.}~\bibnamefont {Salimi}}, \
  and\ \bibinfo {author} {\bibfnamefont {A.~S.}\ \bibnamefont {Khorashad}},\
  }\href {\doibase 10.1038/s41598-022-27156-0} {\bibfield  {journal} {\bibinfo
  {journal} {Scientific Reports}\ }\textbf {\bibinfo {volume} {13}},\ \bibinfo
  {pages} {160} (\bibinfo {year} {2023})}\BibitemShut {NoStop}%
\bibitem [{\citenamefont {Wang}\ and\ \citenamefont {Clerk}(2021)}]{Wang2021}%
  \BibitemOpen
  \bibfield  {author} {\bibinfo {author} {\bibfnamefont {Y.~X.}\ \bibnamefont
  {Wang}}\ and\ \bibinfo {author} {\bibfnamefont {A.~A.}\ \bibnamefont
  {Clerk}},\ }\href {\doibase 10.1038/s41467-021-26868-7} {\bibfield  {journal}
  {\bibinfo  {journal} {Nat. Commun.}\ }\textbf {\bibinfo {volume} {12}},\
  \bibinfo {pages} {6528} (\bibinfo {year} {2021})}\BibitemShut {NoStop}%
\end{thebibliography}%



\appendix

\setcounter{secnumdepth}{3}
\setcounter{equation}{0}
\setcounter{figure}{0}

% Add S in front of figure/equation labels
\renewcommand{\theequation}{S\arabic{equation}}
\renewcommand{\thesubsection}{S\arabic{subsection}}
\renewcommand{\thefigure}{S\arabic{figure}}

\section{Characteristic function of heat and characteristic functions of work arising from the interaction}
\label{app:charfun}
In this appendix we show in detail the calculations that lead to Eq.~\eqref{eq:charFun4} and Eq.~\eqref{eq:charFun_n} in the main text. In the expression for $\Theta (u)$ defined in Eq.~\eqref{eq:charFun}, we use $\hat{U}(t_{f}) = \exp{[-i t_{f} \hat{H}]}$ and substitute the explicit form for $\hat{H}$. Then

\begin{align}
\label{eq:charFunBath}
    \Theta\left(u\right) =& 
     \Tr_{SB}\left[e^{i t_{f} \hat{H}}e^{iu\hat{H}_{B}} e^{-i t_{f} \hat{H}} e^{-iu\hat{H}_{B}}\hat{\rho}\left(0\right)\right] = \nonumber \\
    =& \Tr_{SB}\left[e^{i t_{f}\hat{H}_{B}+it_{f}\hat{H}_{SB}}e^{i t_{f}\hat{H}_{S}}e^{iu\hat{H}_{B}} \times \right. \nonumber \\
    & \left. e^{-it_{f}\hat{H}_{B}-it_{f}\hat{H}_{SB}}e^{-it_{f}\hat{H}_{S}}e^{-iu\hat{H}_{B}}\hat{\rho}_{S}\left(0\right)\otimes\hat{\rho}_{B}\left(0\right)\right],
\end{align}
where we have applied $[\hat{H}_{S},\hat{H}_{SB}]=0$ and the factorized initial state condition. With an interaction Hamiltonian $\hat{H}_{SB} = \sum_{k}g_{k}\hat{\Pi}_{k}\otimes\hat{V}_{k}$, we have
\begin{align}
\label{eq:charFunBath2}
   \Theta\left(u\right) &=     \Tr_{SB}\left[e^{i t_{f}\hat{H}_{B}+i t_{f}\sum_{k}g_{k}\hat{\Pi}_{k}\otimes\hat{V}_{k}}e^{iu\hat{H}_{B}} \times \right. \nonumber \\
    &\left. e^{-i t_{f} \hat{H}_{B}-i t_{f} \sum_{k}g_{k}\hat{\Pi}_{k}\otimes\hat{V}_{k}}e^{-iu\hat{H}_{B}}\hat{\rho}_{S}\left(0\right)\otimes\hat{\rho}_{B}\left(0\right)\right].
\end{align}
The generic open quantum system initial state can be written in the orthonormal eigenbasis of $\hat{H}_S$, $\left\{ \left|n\right\rangle \right\}$, as $\hat{\rho}_{S}\left(0\right) = \sum_{i,j}\rho_{S}^{ij}\left(0\right)\left|i\right\rangle \left\langle j\right|$, where $\rho_{S}^{ij}\left(0\right) = \bra{i}\hat{\rho}_{S}\left(0\right) \ket{j}$. Inserting a resolution of the identity in the eigenbasis of the quantum system energy in Eq.~\eqref{eq:charFunBath2}, and tracing over its degrees of freedom $S$, it is straightforward to calculate
\begin{align}
\Theta\left(u\right) &= \sum_{n}Tr_{B}\left[\sum_{m}\sum_{i}\delta_{nm}e^{it_{f}\hat{H}_{B}+it_{f}g_{n}\delta_{nm}\hat{V}_{n}}e^{iu\hat{H}_{B}} \times \right. \nonumber \\
&\left.\delta_{mi}e^{-it_{f}\hat{H}_{B}-it_{f} g_{m}\delta_{mi}\hat{V}_{m}}e^{-iu\hat{H}_{B}}\rho_{S}^{in}\left(0\right)\hat{\rho}_{B}\left(0\right)\right],\label{eq:charFunBath3}
\end{align}
where we have used $\bra{n}\hat{\Pi}_{k}\ket{m} = \delta_{nk}\delta_{km}$ and $\bra{n}m\rangle = \delta_{nm}$. Using the Kronecker delta properties,  Eq.~\eqref{eq:charFunBath3} simplifies to
\begin{align}
    \Theta\left(u\right) =& \sum_{n}p_{n}Tr_{B}\left[e^{it_{f}\hat{H}_{B}+it_{f}g_{n}\hat{V}_{n}}e^{iu\hat{H}_{B}} \times \right. \\ \nonumber
   &\left. e^{-it_{f}\hat{H}_{B}-it_{f}g_{n}\hat{V}_{n}}e^{-iu\hat{H}_{B}}\hat{\rho}_{B}\left(0\right)\right], \label{eq:charFunBath4}
\end{align}
with $p_{n} = \rho_{S}^{nn}\left(0\right)$. It is straightforward to see from the definition of $\hat{H}_n$ [Eq.~\eqref{H_0_andH_1}] that 
\begin{equation}
    \Theta\left(u\right)=\sum_{n}p_{n}Tr_{B}\left[e^{it_{f}\hat{H}_{n}}e^{iu\hat{H}_{0}}e^{-it_{f}\hat{H}_{n}}e^{-iu\hat{H}_{0}}\hat{\rho}_{B}\left(0\right)\right],
\end{equation}
and that this leads to Eq.~\eqref{eq:charFun4}. A very similar calculation leads to the expression for the time evolution of $\hat{\rho}_B(t)$ defined in the main text. From $\hat{\rho}_B(t) = \Tr_S \left[\hat{U}\left(t\right)\hat{\rho}\left(0\right)\hat{U}^{\dagger}\left(t\right)\right]$, it follows that
\begin{align}
    \hat{\rho}_B(t) &= \sum_{n}\bra{n}e^{-i t\hat{H}_{B}-i t\sum_{k}g_{k}\hat{\Pi}_{k}\otimes\hat{V}_{k}} \times \nonumber \\
    &\sum_{i,j}\rho_{S}^{i,j}\left(0\right)\left|i\right\rangle \left\langle j\right| \hat{\rho}_{B}(0)e^{i t\hat{H}_{B}+i t\sum_{k}g_{k}\hat{\Pi}_{k}\otimes\hat{V}_{k}}\ket{n},
\end{align}
where we have again used the assumption $[\hat{H}_{S},\hat{H}_{SB}]=0$ and the decomposition of the system operator into the eigenbasis of $\hat{H}_{S}$. Using the orthonormality of the open system energy eigenbasis, it follows that 
\begin{align}
    \label{eq:rhoBEv}
    \hat{\rho}_B(t) =& \sum_{n}p_{n}e^{-it\hat{H}_{n}}\hat{\rho}_B\left(0\right)e^{it\hat{H}_{n}}
\end{align}
from which it is clear that $e^{-it\hat{H}_{n}}$ is the operator that defines the time evolution of the bath density matrix.

\section{Decoherence without heat dissipation and its connection to static phase noise}
\label{app:classical_noise}

In this appendix, we show that decoherence processes that do not involve heat absorption by the bath are generically equivalent to classical, static phase noise. From Eqs.~\eqref{eq:charFun4} and~\eqref{eq:charFun_n} we see that a sufficient condition for vanishing heat dissipation is that $[\hat{H}_n,\hat{H}_B] =[\hat{V}_n,\hat{H}_B] =0$ for all $n$. If the Hamiltonian $\hat{H}_B$ is nondegenerate, this further implies that $[\hat{V}_n,\hat{V}_m]=0$ for all $m,n$. That is, the interaction operators $\{\hat{V}_n\}$ share a common eigenbasis with $\hat{H}_B$, i.e., $\hat{V}_n|E_j\rangle = V_n(E_j)|E_j\rangle$, with $V_n(E_j)$ denoting the eigenvalue of $\hat{V}_n$ corresponding to the energy eigenvector $|E_j\rangle$ of $\hat{H}_B$. 

Under these conditions, the overlap function $\langle e^{i\hat{H}_nt} e^{-i \hat{H}_mt}\rangle_B$ entering Eq.~\eqref{reduced_state_matrix_elements} can be written as
\begin{align}
    \label{decoherence_commuting}
    \left \langle e^{i\hat{H}_nt} e^{-i \hat{H}_mt}\right\rangle_B & = \sum_j  p(E_j) e^{i\left(V_n(E_j) - V_m(E_j)\right)t},
\end{align}
which describes the average of a random phase shift that takes the value $V_n(E_j) - V_m(E_j)$ with probability $p(E_j) = e^{-\beta E_j}/Z_B$. This result can be reproduced by a simple noise model in which the dynamical bath is replaced by a random, \textit{time-independent} Hamiltonian
\begin{equation}
    \label{H_random}
    \hat{H}_S = \sum_n (\varepsilon_n + \delta_n) |n\rangle\langle n|,
\end{equation}
where the energy shift $\delta_n$ is a stochastic variable that takes the possible values $V_n(E_j)$ with corresponding probabilities $p(E_j)$. It is straightforward to verify that the ensemble-averaged state $\hat{\rho}_S(t) = \langle \!\langle e^{-i\hat{H}_S t} \hat{\rho}_S(0) e^{i\hat{H}_S t} \rangle\!\rangle$ has matrix elements given by Eqs.~\eqref{reduced_state_matrix_elements} and \eqref{decoherence_commuting}, where $\langle \!\langle \bullet \rangle\!\rangle$ denotes an average over the noise distribution $p(E_j)$. We have therefore shown that the absence of heat dissipation implies that the dephasing noise is equivalent to a \textit{static}, random phase shift for each eigenstate of the open quantum system. Such noise is characterised by a spectrum with vanishing weight away from zero frequency, and can be distinguished from time-dependent (quantum or classical) phase noise by standard noise spectroscopy techniques~\cite{Alvarez2011,Norris2016}.

We note that, since $[\hat{V}_n,\hat{H}_B] =0$ for all $n$ is only a sufficient condition, we cannot rule out the absence of heat dissipation when $[\hat{V}_n,\hat{H}_B] \neq 0$ for certain initial environment states, e.g., infinite-temperature states. Furthermore, the arguments above have assumed that $\hat{H}_B$ is nondegenerate. If we allow $\hat{H}_B$ to be degenerate, it is possible to construct models in which $[\hat{V}_n,\hat{H}_B] =0$ for all $n$ yet $[\hat{V}_n,\hat{V}_m] \neq 0$ for some $m,n$ (e.g., $\hat{H}_B = \hat{\mathbf{L}}^2$, $\hat{V}_1 = \hat{L}_x$, and $\hat{V}_2 = \hat{L}_y$, with $\hat{\mathbf{L}} = (\hat{L}_x,\hat{L}_y,\hat{L}_z)$ the angular momentum). Remarkably, such models can feature nontrivial decoherence for the open system despite the state of the bath being completely invariant under the dynamics.

\section{Functional determinant approach}
\label{app:levitov}

In this appendix, we discuss the method used to exactly calculate Eq.~\eqref{eq:ThetaModel} for a bath of fermionic modes. The functional determinant approach for fermionic Hamiltonians states that the mean value of the product of a finite number of exponential fermionic operators can be expressed using the Levitov formula~\cite{Klich2003},
\begin{equation}
    \left\langle e^{\hat{Y}_{1}}e^{\hat{Y}_{2}}...e^{\hat{Y}_{N}}\right\rangle _{B} = \det\left[1-\hat{n}+\hat{n}e^{\hat{y}_{1}}e^{\hat{y}_{2}}...e^{\hat{y}_{N}}\right] \label{eq:LevitovFormula},
\end{equation}
where $\hat{Y}_{i}$ are bilinear many-body fermionic operators, $\hat{y}_{i}$ are their corresponding single-particle operators, and $\hat{n}$ is the single-particle occupation number operator. When applied to the second term of Eq.~\eqref{eq:ThetaModel}, the Levitov formula yields
\begin{align}
\label{eq:charFunLevitov}
&\Theta_1 \left(u\right)=
\det\left[1-\hat{n}+\hat{n} e^{it_f\hat{h}_{1}}e^{i u\hat{h}_{0}} e^{-it_f\hat{h}_{1}}e^{-i u\hat{h}_{0}}\right].
\end{align}
We use this to exactly calculate the characteristic function for the model discussed in the main text.

\clearpage
\end{document}